\documentclass[preprint]{aastex}

\shorttitle{Microflares in Canopy-type Configuration}
\shortauthors{Jiang, R.-L. et al.}

\begin{document}

\title{Numerical Simulation of Solar Microflares in a Canopy-Type Magnetic Configuration}

\author{R.-L. Jiang, C. Fang, and P.-F. Chen}

\affil{School of Astronomy and Space Science, Nanjing University, Nanjing 210093, China
\and Key Laboratory of Modern Astronomy and Astrophysics (Nanjing University), Ministry of Education, China}
\email{Email: rljiang@nju.edu.cn}

\newcommand{\figurepath}{.}

\begin{abstract}
Microflares are small activities in solar low atmosphere, some are in the low 
corona, and others in the chromosphere. Observations show that some of the 
microflares are triggered by magnetic reconnection between emerging flux and a 
pre-existing background magnetic field. We perform 2.5D compressible resistive 
MHD simulations of magnetic reconnection with gravity considered. 
The background magnetic field is a canopy-type configuration which is rooted 
at the boundary of the solar supergranule. By changing the bottom boundary 
conditions in the simulation, new magnetic flux emerges up at the center of 
the supergranule and reconnects with the canopy-type magnetic field. We 
successfully simulate the coronal and chromospheric microflares, whose current 
sheets are located at the corona and the chromosphere, respectively. The 
microflare of coronal origin has a bigger size and a higher temperature 
enhancement than that of chromospheric origin. In the microflares of coronal 
origin, we also found a hot jet ($\sim$$1.8 \times 10^6$ K), which is 
probably related to the observational EUV/SXR jets, and a cold jet 
($\sim$$10^4$ K), which is similar to the observational H$\alpha$/Ca surges, 
whereas there is only an H$\alpha$/Ca bright point in the microflares of 
chromospheric origin. The study of parameter dependence shows that the 
size and strength of the emerging magnetic flux are the key parameters 
which determine the height of the reconnection location, and further determine 
the different observational features of the microflares.

\end{abstract}


\section{INTRODUCTION}

Microflares, or subflares, which are small-scale and short-lived solar 
activities, have already been studied for many years since last century
~\citep{Smith1963, Svestka1976, Tandberg-Hanssen1988}. The typical size, 
duration and total released energy are 5--20 arcsecs, 10--30 minutes and
$10^{26}$--$10^{29}$ ergs \citep{Shimizu2002, Fang2006, Fang2010}, 
respectively. Microflares have been observed in many wavelengthes 
including H$\alpha$ \citep{Schmieder1997, Chae1999, Tang2000}, EUV 
\citep{Emslie1978, Porter1984, Chae1999, Brosius2009, Chen2010}, 
soft X-ray \citep{Golub1974, Golub1977, Tang2000, Shimizu2002, Kano2010},
hard X-ray \citep{Lin1984, Qiu2004, Ning2008, Brosius2009}, and microwave 
\citep{Gary1988, Gopalswamy1994, Gary1997}. However, not all microflares 
have emissions at all wavelengths. Observations show that most of bright 
X-ray microflares also appear at EUV and H$\alpha$ bands, but only part of 
H$\alpha$ microflares have their counterparts in X-ray emission 
\citep{Zhang2012}.

Observational characteristics of microflares, such as the heating, the 
relation with magnetic field, the duration and the coincidence between 
different wavelengths, etc, imply that microflares are produced by 
magnetic reconnection, similar to big flares. For example, \cite{Qiu2004} 
found that about 40\% of microflares show hard X-ray emissions at over 10 
keV and microwave emissions at about 10 GHz, typical features for flares.
Recently, \cite{Ning2008} found that roughly half of the microflares display
the Neupert effect as previously revealed in flares. On the other hand, some
other microflares may have their origins in the lower atmosphere. 
\cite{Brosius2009} found that the microflares are bright in the chromospheric 
and transition region spectral lines, which is consistent with the 
chromospheric heating by nonthermal electron beams. \cite{Jess2010} concluded 
that the microflares in their study are due to magnetic reconnection at a
height of 200 km above the solar surface.

Since a part of the microflares are located at emerging flux regions, they
are probably due to magnetic reconnection driven by the new emerging magnetic 
flux (EMF) with a pre-existing magnetic field. \cite{Schmieder1997} found that 
the X-ray loops of microflares appear at the locations of EMF. \cite{Chae1999} 
found some repeatedly occurred EUV jets where the pre-existing magnetic 
field was ``canceled" by the new EMF with opposite polarity. \cite{Tang2000} 
indicated that the new EMF successfully emerged about 20 minutes before the 
peaks of H$\alpha$ and soft X-ray brightenings. Half of the events studied 
by \cite{Shimizu2002} show the small scale emergences of magnetic flux loops 
in the vicinity of the transient brightenings. \cite{Kano2010} found that 
EMFs and moving magnetic features (MMFs) are related to the energy release 
for at least half of the microflares around a well developed sunspot.
Therefore, to understand the dynamics of microflares, it is crucial to simulate
the magnetic reconnection associated with EMF.

The numerical simulations of flux emergence scenario started from 1980s 
\citep{Horiuchi1988, Matsumoto1988, Shibata1989a, Shibata1989b}. Later, more 
realistic and complicated numerical experiments \citep{Magara2001, Fan2001, 
Manchester2004, Isobe2005, Leake2006, Isobe2007, Hood2009} were presented to study the 
emergence of twisted flux from the convection region to solar corona. The 
simulations are mainly based on the Parker instability which is proposed by 
\cite{Parker1966} and their results can be applied to account for the formation 
of filament or newly active region. More 2D and 3D numerical experiments are 
focused on the interaction between the emerging flux and the per-existing 
coronal magnetic configuration to study the solar eruptions. With 2D 
simulations, it is found that the magnetic reconnection due to flux emergence 
can well explain solar jets or explosive events \citep{Yokoyama1995, Jin1996, 
Nishizuka2008, Ding2010, Ding2011}, or can even trigger the onset of coronal mass
ejections \citep{Chen2000}. Further 3D simulations gave the same but more
detailed results, which revealed that the emerging flux plays a key role in 
the formation of many solar activities, such as type II spicules, erupting 
filaments or flux ropes, Ellerman bombs, and so on \citep{Galsgaard2005, 
Galsgaard2007, Torok2009, Archontis2008, Archontis2009, Mart2011}. 

From the description of previous works, it is seen that EMF is responsible for
many different phenomena, and one main cause of the diversity of the dynamics
is that the reconnection happens at diffferent heights in the solar atmosphere
\citep{Shibata1996, Chen1999}. Our simulation in this paper is mainly focused 
on the microflares in a canopy-type magnetic configuration, where the height of
the reconnection is self-consistently determined by the EMF. In our previous 
papers \citep{Jiang2010, Xu2011}, we presented the 2.5 dimensional 
(2.5D) magnetohydrodynamic (MHD) simulation of magnetic reconnection in the 
chromosphere, which reproduces qualitatively the temperature enhancement 
observed in chromospheric microflares. However, the magnetic configuration
in that simulation, the same as in \cite{Chen2001}, is too idealized and the 
reconnection site was specified at the computational center. Besides, no corona was included, so 
that the simulations can tell nothing about the EUV and SXR features. In this 
paper, we adopt a more realistic magnetic configuration and extend the solar
atmosphere up to the corona in order to see what determines a microflare to be
of coronal origin or of the chromospheric origin, and to compare the
difference between the two types of microflares. The paper is organized as 
follows: the numerical method is described in Section \ref{numerical_method}; 
the numerical results are presented in Section \ref{Sec:Numerical_Results}. 
Discussion and summary are given in Section \ref{Sec:Discussion_Summary}.

\section{NUMBERICAL METHOD}
\label{numerical_method}

\subsection{Basic Equations}
How to keep the magnetic field divergence free is one of the big problems in 
all MHD codes. Because of the discretization and numerical errors, the 
performance of the MHD code can be unphysical \citep{Brackbill1980}. There are several ways to maintain $\nabla
\cdot \mathbf{B} = 0$ for MHD equations: (1) 8-wave formulation 
\citep{Powell1999}, (2) the CT method \citep{Evans1988, Stone1992}, (3) the projection 
scheme \citep{Brackbill1980}. The comparison between these methods showed that 
different methods have their own advantages and disadvantages \citep{Toth2000}. 
Besides, one can rewrite the original MHD equations by using 
\itshape vector potential \upshape $\mathbf{A}$ instead of the magnetic field
$\mathbf{B}$, or by using the vector magnetic potential or Euler potential. 
The advantage is that the divergence free condition is always satisfied, however, the MHD
equation should be rewritten. Our method, as introduced in \cite{Jiang2012} 
adopts the 8-wave method. As suggested by \cite{Dedner2002}, we use the Extended 
Generalized Lagrange Multiplier (EGLM)-MHD equations rather than pure 
MHD equations, which include two additional waves 
to transfer the numerical error of $\nabla \cdot \mathbf{B}$. The local 
divergence error can be damped and passed out of the computational domain. 
The adopted dimensionless EGLM-MHD 
equations with resistivity and gravity included are given as follows:

\begin{equation}
\frac{\partial \rho}{\partial t} + \nabla \cdot \left(\rho \mathbf{v}\right)
= 0 \,\, , \label{Equ:MHD-1}
\end{equation}

\begin{equation}
\frac{\partial \left(\rho \mathbf{v}\right)}{\partial t} + \nabla \cdot
\left( \left(p + \frac{1}{2}{B}^2 \right) \mathbf{I} + \rho \mathbf{v}
\mathbf{v} - \mathbf{B} \mathbf{B}\right) =  - \left(\nabla \cdot \mathbf{B}
\right) \mathbf{B} + \rho \mathbf{g} \,\, , \label{Equ:MHD-2}
\end{equation}

\begin{equation}
 \frac{\partial \mathbf{B}}{\partial t} + \nabla \cdot \left(\mathbf{v}
 \mathbf{B} - \mathbf{B} \mathbf{v} + \psi \mathbf{I} \right) = - \nabla \times
 \left(\eta \nabla \times \mathbf{B}\right) \,\, , \label{Equ:MHD-3}
\end{equation}

\begin{equation}
\frac{\partial e}{\partial t} + \nabla \cdot \left(\mathbf{v}
\left(e + \frac{1}{2}{B}^2 + p\right) - \mathbf{B} \left(\mathbf{B}
\cdot \mathbf{v} \right)\right) = - \mathbf{B} \cdot \left(\nabla
\psi \right) - \nabla \cdot \left( \left(\eta \nabla \times
\mathbf{B} \right) \times \mathbf{B} \right) + \rho \mathbf{g} \cdot \mathbf{v} \,\, ,
\label{Equ:MHD-4}
\end{equation}

\begin{equation}
\frac{\partial \psi}{\partial t} + c_h^2 \nabla \cdot \mathbf{B} =
-\frac{c_h^2}{c_p^2} \psi \,\, , \label{Equ:MHD-5}
\end{equation}

\noindent 
where eight independent conserved variables are the density ($\rho$), momentum 
($\rho v_x$, $\rho v_y$, $\rho v_z$), magnetic field ($B_{x}$, $B_{y}$, 
$B_{z}$), and total energy density ($e$). The expression of the total energy 
density is $e = p/(\gamma-1) + \rho v^2/2 + B^2/2$ where $\gamma = 1.1$ is 
taken in all our computations. The pressure $p$ and temperature
$T$ are dependent on the eight conserved variables, $\mathbf{g}$ is
the gravity vector, and $\eta$ the magnetic resistivity coefficient. Finally, a unity matrix
$\mathbf{I}$ is the unity matrix. The main variables are
normalized by the quantities given in Table \ref{Table:Normalization}.

\begin{table}[t]
\centering
\caption{Normalization Units}
\label{Table:Normalization}
\begin{tabular}{cccc}
\hline
\hline
Variable & Quantity & Unit & Value \\
\hline
$T$          & Temperature    & $T_0$                                         & $10000$ K                            \\
$\rho$       & Density        & $\rho_0$                                      & $2.56 \times 10 ^ {-8}$ kg m$^{-3}$  \\
$x,y$        & Length         & $L_0=T_0 \kappa_B / (m g_s)$                  & $301.6$ km                           \\
$p$          & Pressure       & $p_0=\rho_0 T_0 \kappa_B / m$                 & $2.12$ N m$^{-2}$                    \\
$\mathbf{V}$ & Velocity       & $v_0=(p_0 / \rho_0)^{1/2}$                    & $9.09$  km s$^{-1}$                  \\
$\mathbf{B}$ & Magnetic field & $B_0=(\mu_0 p_0)^{1/2}$                       & $16.3$ G                             \\
$t$          & Time           & $t_0=L_0 / v_0$                               & $33.1 $ s                            \\
\hline
\end{tabular}
\end{table}

In the EGLM-MHD equations, $\psi$ is a scalar potential propagating the 
divergence error, $c_h$ is the wave speed, and $c_p$ the damping rate of 
the wave \citep{Dedner2002, Matsumoto2007}. As suggested by \cite{Dedner2002}, 
the expressions for $c_h$ and $c_p$ are:

\begin{equation}
c_h = \frac{c_{cfl}} {\Delta t} \min (\Delta x, \Delta y) \,\, ,
\end{equation}

\begin{equation}
c_p = \sqrt{-\Delta t \frac{c_h^2} {\ln{c_d}}} \,\, ,
\end{equation}

\noindent
where $\Delta t$ is the time step, $\Delta x$ and $\Delta y$ 
are the grid sizes, $c_{cfl}$ is a safety coefficient less than
1. $c_d \in (0, 1)$ is a problem dependent coefficient to decide the
damping rate for the waves of divergence errors. We can see that
$c_h$ and $c_p$ are not independent of the grid resolution and the
scheme used. Hence we have to adjust their values for different
situations.

\subsection{Initial Condition}
\label{InitialCondition}

\begin{figure}[!htbp]
   \centering
   \includegraphics[width=300pt]{\figurepath/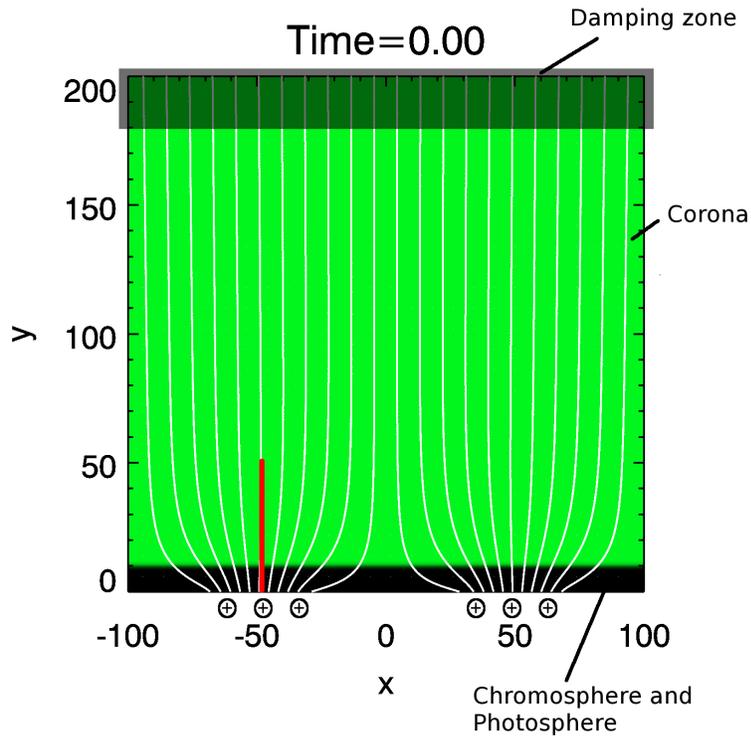}
   \caption{Initial condition for the magnetic configuration and temperature 
   distribution. The ranges of the computational box is from $-100$ to $100$ 
   in the $x$-direction and from $0$ to $200$ in the $y$-direction. The 
   shadowy region means the damping zone. The ``magnetic charges" 
   are located beneath the solar surface. Note that the 
   ``magnetic charges" are samples to show the positions of 
   these ``charges". The units of length, temperature, time 
   and velocity are $301.6$ km, $10000$ K, $33.1 $ s and $9.09$ km s$^{-1}$, respectively.
   \label{fig01}}
\end{figure}

\begin{figure}[!htbp]
   \centering
   \includegraphics[width=300pt]{\figurepath/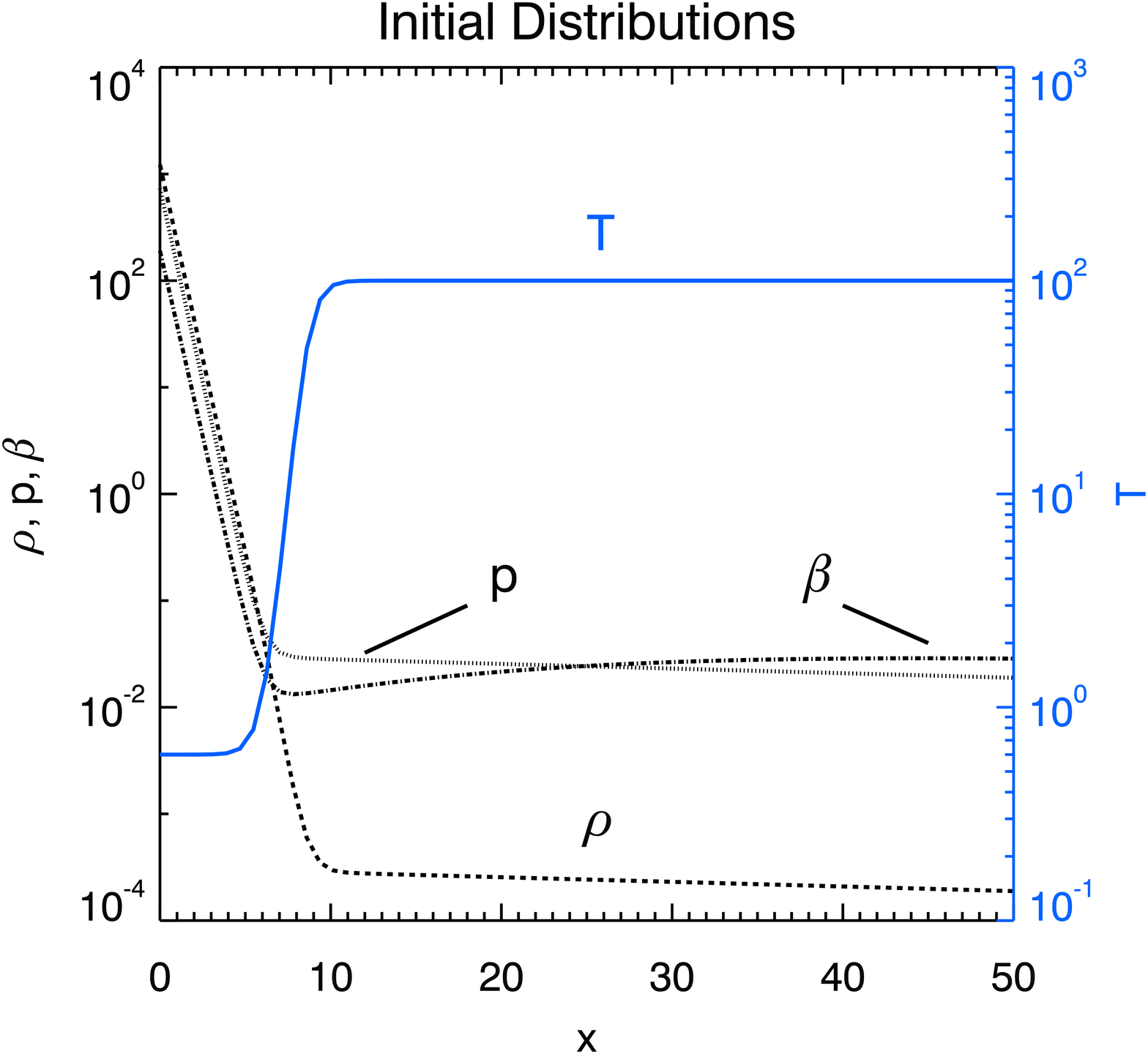}
   \caption{Initial distributions of the density ($\rho$), gas pressure ($p$), 
   plasma beta ($\beta$) and temperature ($T$). The density, gas pressure and plasma $\beta$ distributions
   use the same left $y$-axis and the right blue $y$-axis is for the temperature.
   The units of density, gas pressure, length, temperature, time 
   and velocity are $2.56 \times 10 ^ {-8}$ kg m$^{-3}$, $2.12$ N m$^{-2}$, $301.6$ km, 
   $10000$ K, $33.1 $ s and $9.09$ km s$^{-1}$, respectively.
   \label{fig02}}
\end{figure}

The computational box is located in the $x$-$y$ Cartesian plane. The $x$-axis 
is parallel to the solar surface while the $y$-axis is perpendicular to the 
photosphere. As shown in Figure~\ref{fig01}, the computational domain is 
$-100\leq x \leq 100$ and $0\leq y \leq 200$, where the length unit is
$L_0=301.6$ km. In the simplified chromosphere and photosphere the temperature 
is set uniform with the value of 0.6, after being normalized by $T_0=10000$ K. 
This layer extends from the bottom to $y=8.5$. After a thin transition
region with a thickness of 1, the plasma temperature rises rapidly to 100 in
the corona, which corresponds to 1 MK. Given the temperature distribution we 
can get the density and pressure distributions according to the hydrostatic 
equilibrium. The initial distributions of density, gas pressure, plasma 
$\beta$ (the ratio of gas to magnetic pressures) and temperature along 
the red line in Figure \ref{fig01} are outlined by Figure \ref{fig02}.

As shown in Figure~\ref{fig01}, two canopy-shaped magnetic structures are
separated by a distance of 100 (corresponding to 30 Mm) in order to mimic
the two boundaries of a supergranular cell. The canopy magnetic field, which 
was originally proposed by \cite{Gabriel1976} and \cite{Giovanelli1980, 
Giovanelli1982} for studying the chromospheric and photospheric magnetic 
fields, is rooted at the supergranular boundaries. In order to create a canopy 
magnetic field, we introduce several ``magnetic charges"
locating beneath the photosphere. Accroding to the Gauss's theorem in the
magnetism, the magnetic field (at the position $\mathbf{r}$) generated by a 
``magnetic charges" (at the position $\mathbf{r'}$) can be 
written as $\mathbf{B} = c_B (\mathbf{r} - \mathbf{r'}) / |\mathbf{r} - 
\mathbf{r'}|^2 $ in two dimensional geometry ($c_B$ is a constant related the 
strength of the field) and $\mathbf{B} = c_B (\mathbf{r} - 
\mathbf{r'}) / |\mathbf{r} - \mathbf{r'}|^3 $ in three dimensional geometry, which is 
similar to the electric field. The locations of the ``magnetic charges" are
depicted in Figure~\ref{fig01}. The magnetic configuration
in our simulation is not exactly the same as the canopy model proposed by 
\cite{Gabriel1976} and \cite{Giovanelli1980, Giovanelli1982}.

\subsection{Resistivity Model}
In order to reproduce a fast reconnection~\citep{Petshek1964}, we adopt 
a non-uniform anomalous resistivity. This kind of resistivity may be 
due to some microscopic instabilities, although how these instabilities 
drive the macroscopic reconnection is still not clear. The mathematical 
form of the assumed anomalous resistivity model \citep{Ugai1985, Chen2000, 
Yokoyama2001} is:

\begin{equation}
   \eta = \min \left( \left( \frac{|\mathbf{j}|} {j_c} - 1 \right), \eta_{max} \right) \ for \ |j| \ge j_c \,\, ,
   \label{Equ:Resistivity}
\end{equation}

\noindent
where $|\mathbf{j}|$ is the total current density, $j_c$ the threshold, 
above which the anomalous resistivity is excited, and $\eta_{max}$ the maximum 
value of the resistivity. As shown by Equations (\ref{Equ:Resistivity}), 
the resistivity ($\eta$) is dependent on the current density. The anomalous 
resistivity ($\eta$) is switched off when $|\mathbf{j}| < j_c$.

\subsection{Boundary condition}
The magnetic flux emergence is realized by changing conditions at the lower boundary 
\citep{Forbes1984, Chen2000, Ding2010}. In our case, $B_x$ and $B_y$, rather
than the magnetic flux function in previous works, are independent variables,
we change directly the magnetic field value at the boundary. The magnetic 
field ($B_x$, $B_y$, $B_z$) changes with time while the other variables, e.g., 
$\rho$, $\mathbf{v}$, and $p$, are fixed to the initial values at this 
boundary. The mathematic form of the EMF is given by:

\begin{equation}
B_x(x, y) = -B_e \frac{t}{t_e} \frac{y - y_e}{x^2+y^2} \,\, ,
\label{Equ:EMF01}
\end{equation}
\begin{equation}
B_y(x, y) =  B_e \frac{t}{t_e} \frac{x - x_e}{x^2+y^2} \,\, ,
\label{Equ:EMF02}
\end{equation}

\noindent
in the range $x^2+y^2 \le r_0^2$, $y \le 0$, where $B_e$ is the strength of the EMF, $t_e$ the end time of 
emergence, $r_0$ the half width of the EMF, $x_e$ and $y_e$ the center 
coordinates of the EMF in $x$- and $y$-direction. In our simulation cases, 
$x_e$ and $y_e$ are set to be $0$ and $-3.6$, respectively. 
Formulae (\ref{Equ:EMF01})--(\ref{Equ:EMF02}) are applied to the magnetic 
field $\mathbf{B}$ until $t = t_e$. Later, all the physical values are fixed.

Free boundaries (equivalent extrapolation) are adopted at the upper side 
$y = 200$. However, because of the low plasma beta ($\beta = 2p / \mathbf{B} 
^ 2$) the upper boundary becomes unstable when the jets or waves propagate to 
this boundary. Thus we add a damping zone from $y = 180$ to $y = 200$ in 
$y$-direction as shown in Figure~\ref{fig01}. The waves or jets will be damped 
to the initial value as the formulae given below:

\begin{equation}
   u(x, y) = u(x, y) D(y) + u_0(x, y) \left(1 - D(y) \right) \,\, ,
\end{equation}
\begin{equation}
   D(y) = \frac {1} {2} \left(1 - \tanh \left(\frac{6}{y_{e} - y_{s}} 
\left(y - \frac{y_{s} + y_{e}}{2} \right) \right) \right) \,\, ,
\end{equation}

\noindent
where $u$ represents varibles $\rho$, $\mathbf{v}$, and $p$ (not 
$\mathbf{B}$), $u_0$ is the initial value, $D$ the damping function, $y_s$ the 
start of position for the damping zone in $y$ direction, $y_e$ the end 
position for damping zone. In all our cases, $y_s = 180$ and $y_e = 200$. 
Nonetheless, the damping zone may also lead to some reflections of waves or 
jets, but these reflections have little effect on the results since they are 
far away from the lower atmosphere which we are interested in. We use fixed
boundary condition for the left and right boundaries at $x = -100$ and 
$x = 100$. 

\subsection{Numerical Scheme and AMR Grid}
The code used in our simulation is MAP \citep{Jiang2012}, which is developed 
by the solar group of Nanjing University. The MAP code is a FORTRAN code for 
MHD calculation with the adaptive mesh refinement (AMR) \citep{Berger1984, Berger1989} 
and Message Passing Interface (MPI) parallelization. MAP have three optional 
numerical schemes for the MHD part, namely, modified Mac Cormack Scheme \citep{Yu2001}, Lax-Fridrichs scheme 
\citep{Toth1996} and weighted essentially non-oscillatory (WENO) \citep{Jiang1999} scheme. 
In this paper, we mainly use the WENO scheme, which has a higher accuracy than 
the other two. Moreover, it can keep the total variation diminishing 
(TVD) \citep{Harten1997} property for the MHD part without any 
additional artificial viscosity. The base resolution of our simulation is 
$256 \times 256$ and the mesh refinement level is $4$ in all of our cases. Thus the effective 
resolution is $2048 \times 2048$ globally and the minimum grid size is $0.1$ (about $30$ km).

\section{RESULTS}
\label{Sec:Numerical_Results}
Our simulations show two typical cases, one with the reconnection occurring
in the corona while the other mainly in the chromosphere and partly in transition region. 
The dynamics of magnetic 
reconnection processes is similar in the early stage in the two cases. Both
show the following evolution: (1) an EMF 
emerges from the center of the supergranule; (2) the EMF reconnects with the pre-existed magnetic field; (3) 
temperature is enhanced near the X-point and the reconnection inflow and 
outflow appear. However, the final results are very different. 
The difference between these two cases are presented
in \ref{Sec:Corona_Case} and \ref{Sec:Chromosphere_Case} subsections.

\subsection{Coronal Microflare}
\label{Sec:Corona_Case}

\begin{figure}[!htbp]
   \centering
   \includegraphics[width=300pt]{\figurepath/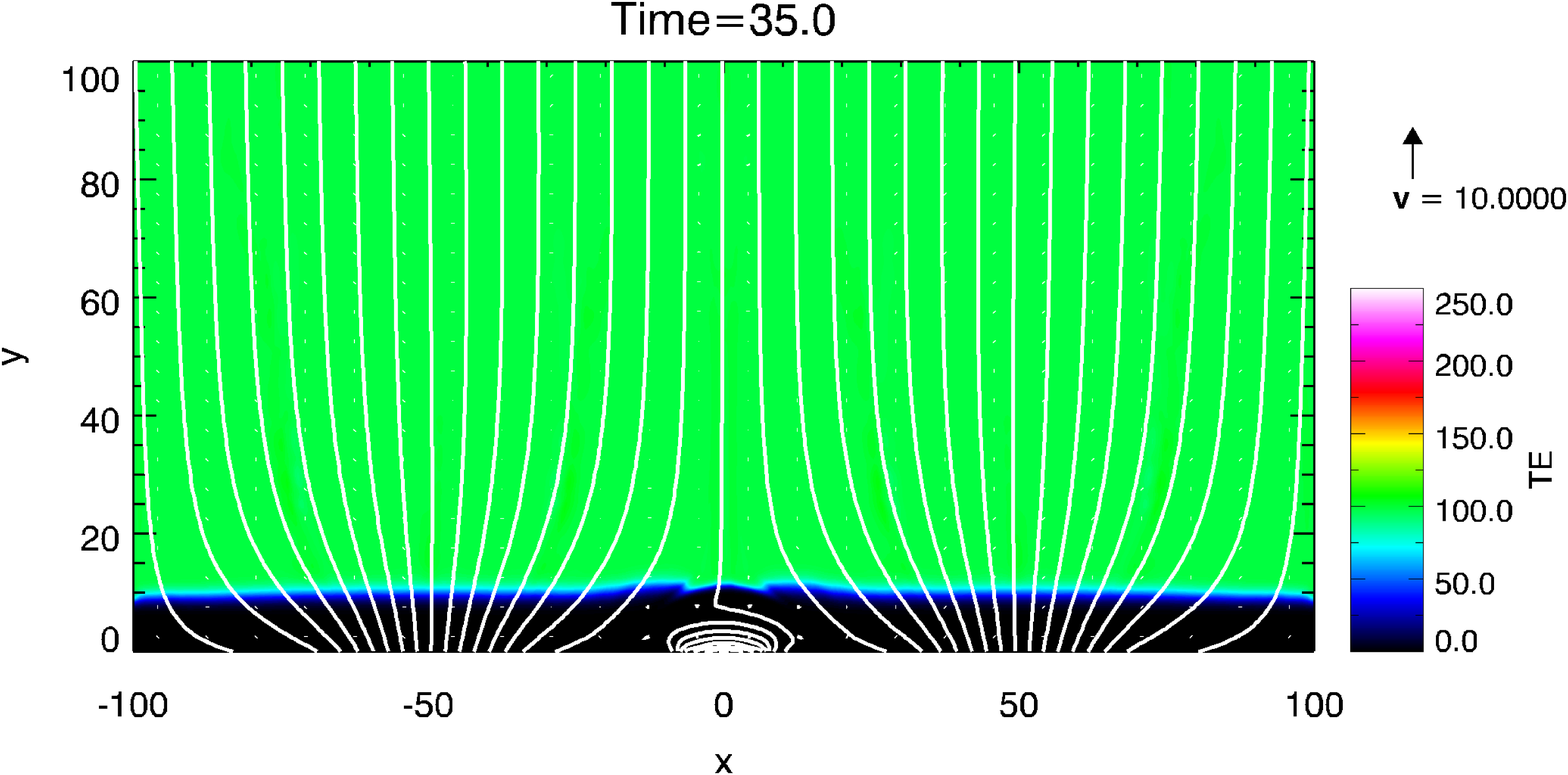}
   \includegraphics[width=300pt]{\figurepath/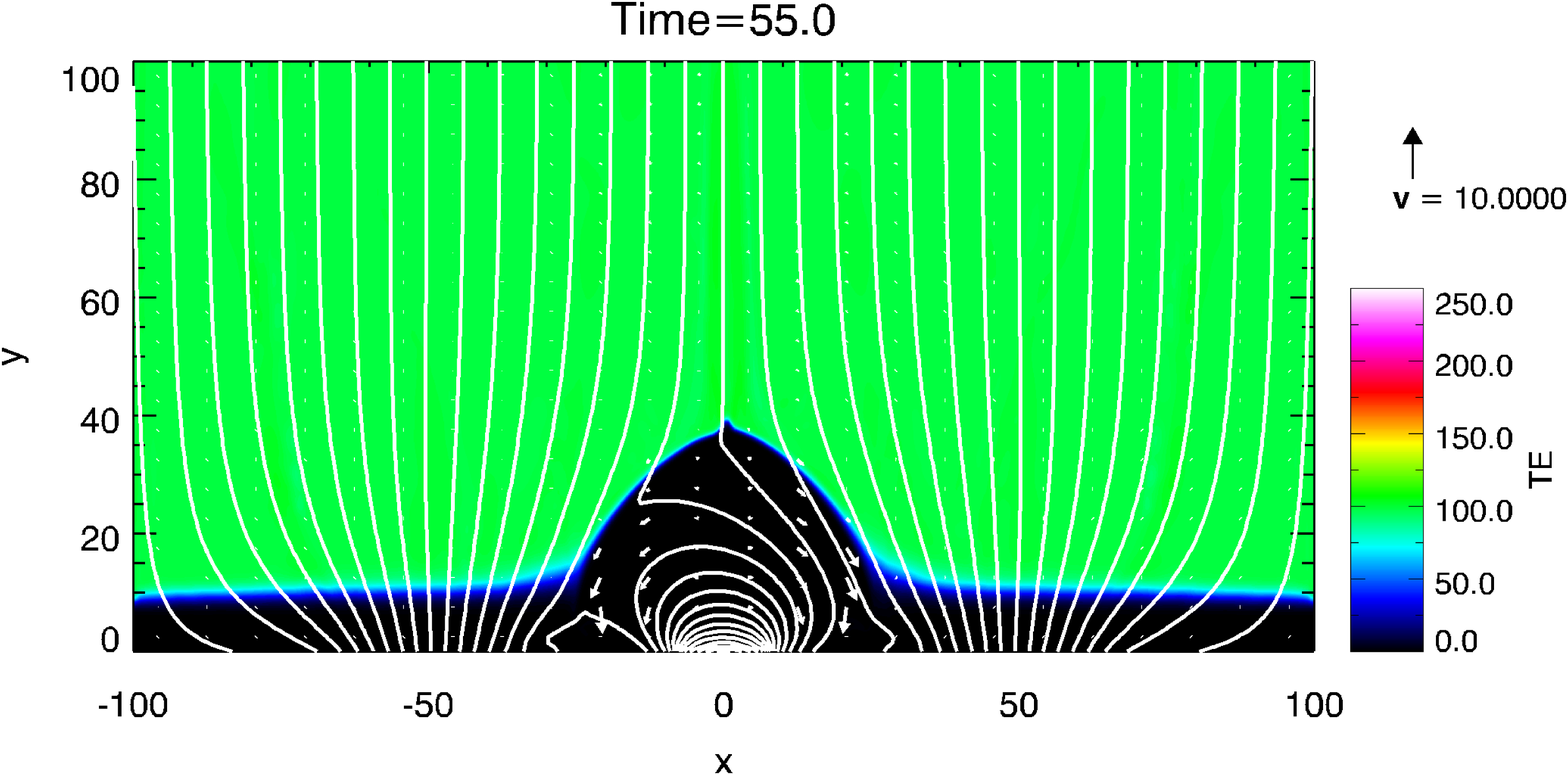}
   \includegraphics[width=300pt]{\figurepath/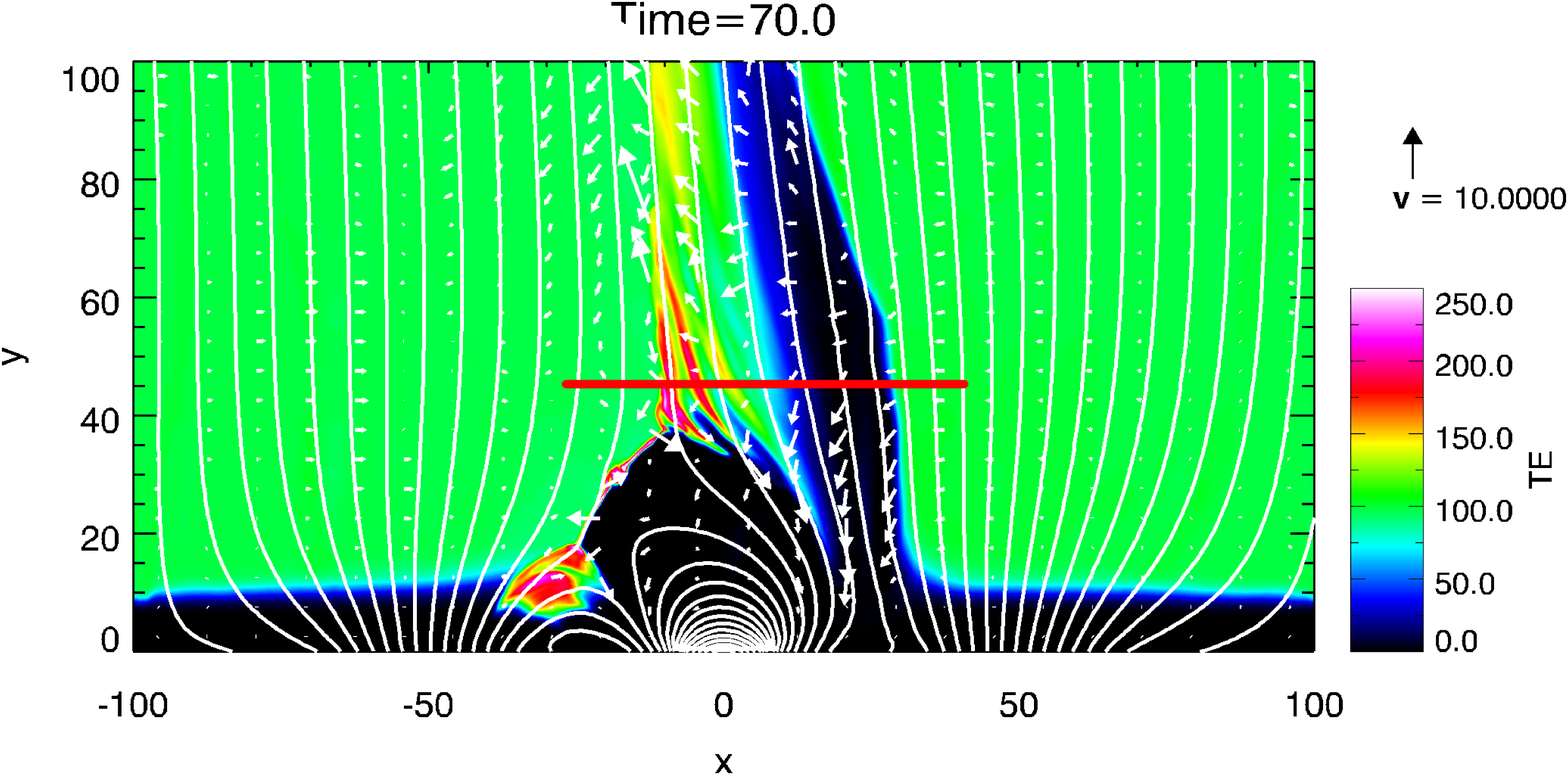}
   \caption{Temperature distributions (color scale), projected 
   magnetic field (solid lines) and velocity field (vector arrows) at
   times 35 (upper panel), 55 (middle panel) and 70 (lower panel).
   The units of length, temperature, time and velocity are $301.6$ km, 
   $10000$ K, $33.1 $ s, $9.09$ km s$^{-1}$, respectively.
   \label{fig03}}
\end{figure}

\begin{figure}[!htbp]
   \centering
   \includegraphics[width=200pt]{\figurepath/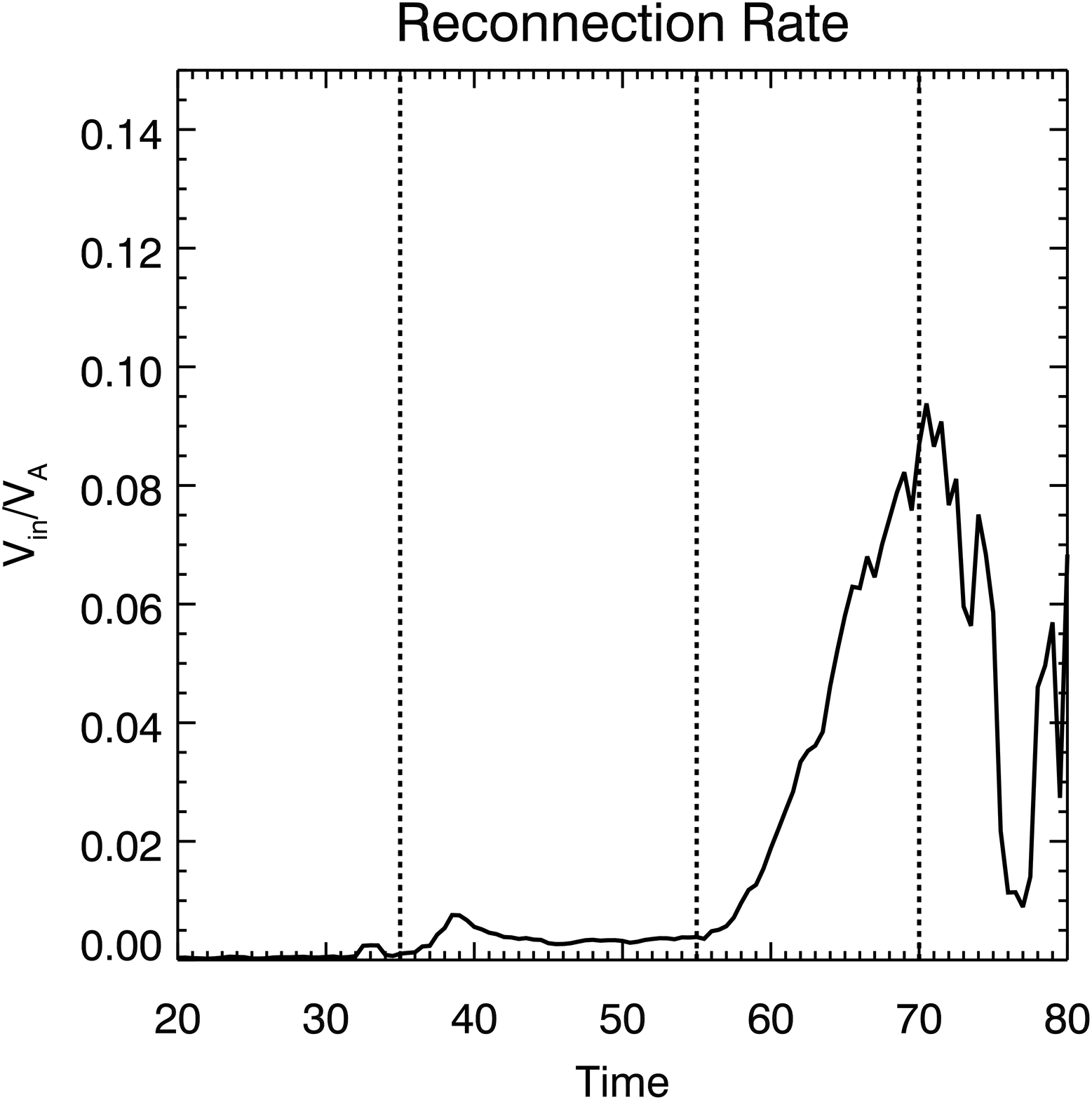}
   \includegraphics[width=200pt]{\figurepath/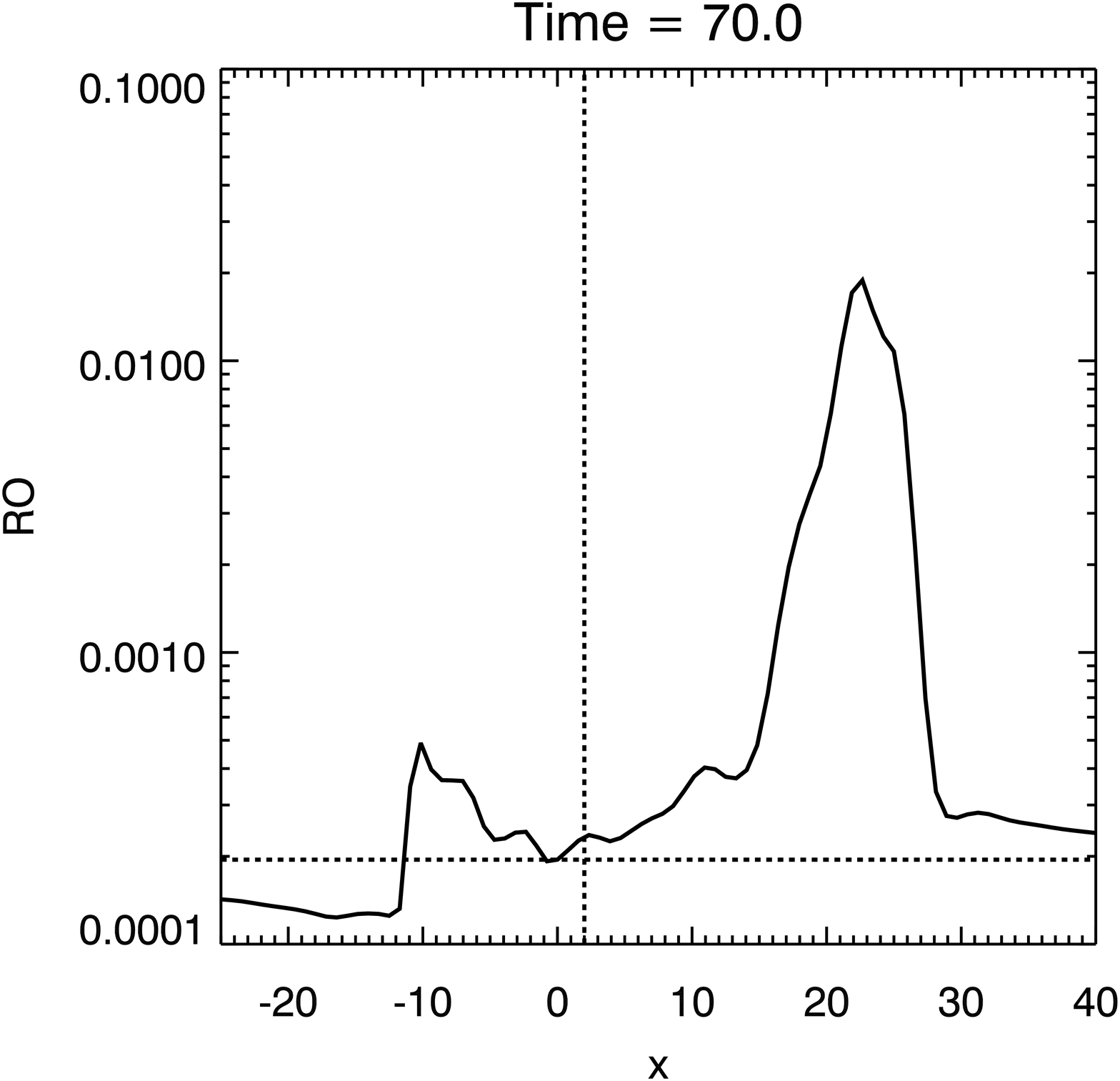}
   \includegraphics[width=200pt]{\figurepath/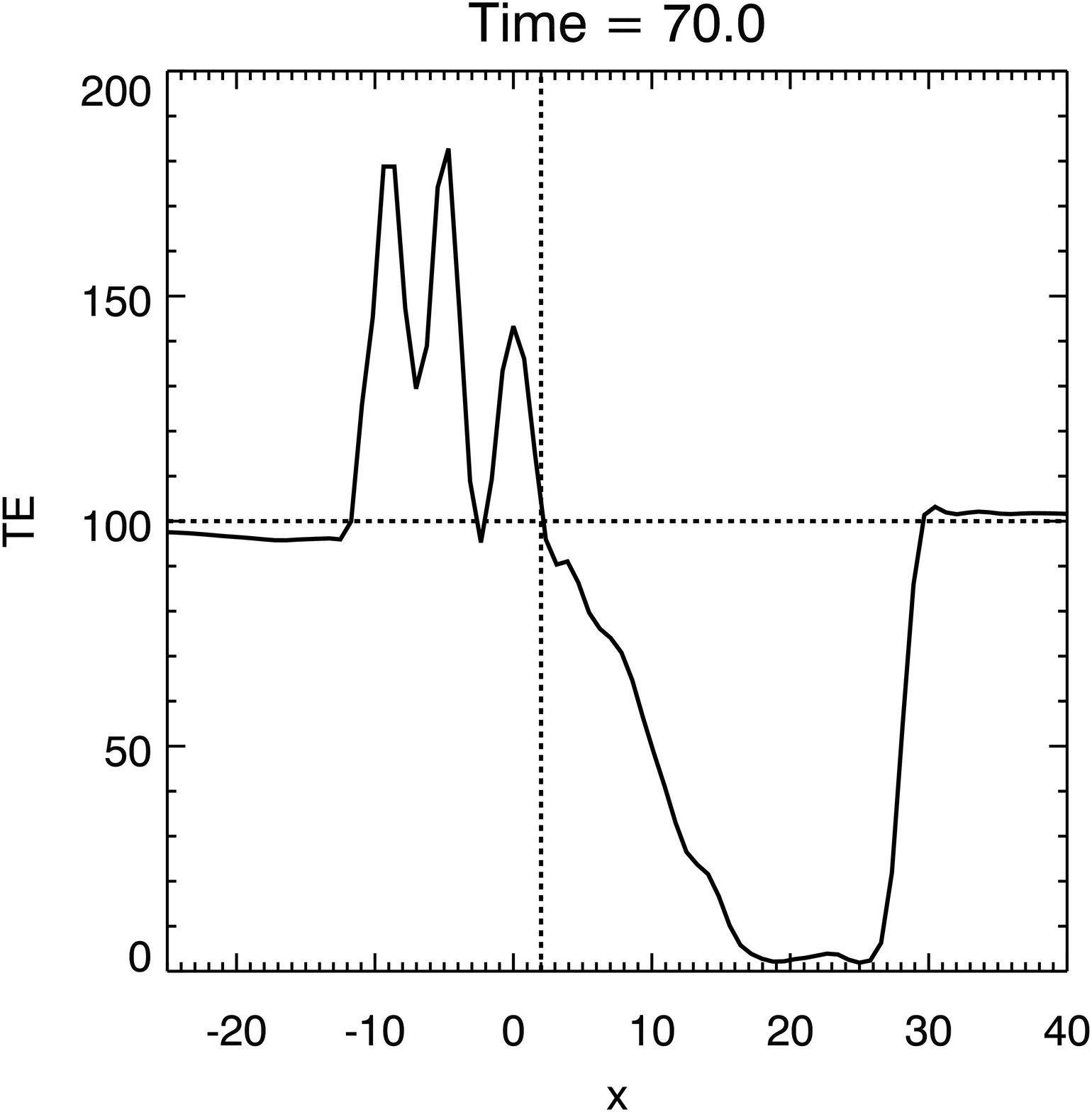}
   \includegraphics[width=200pt]{\figurepath/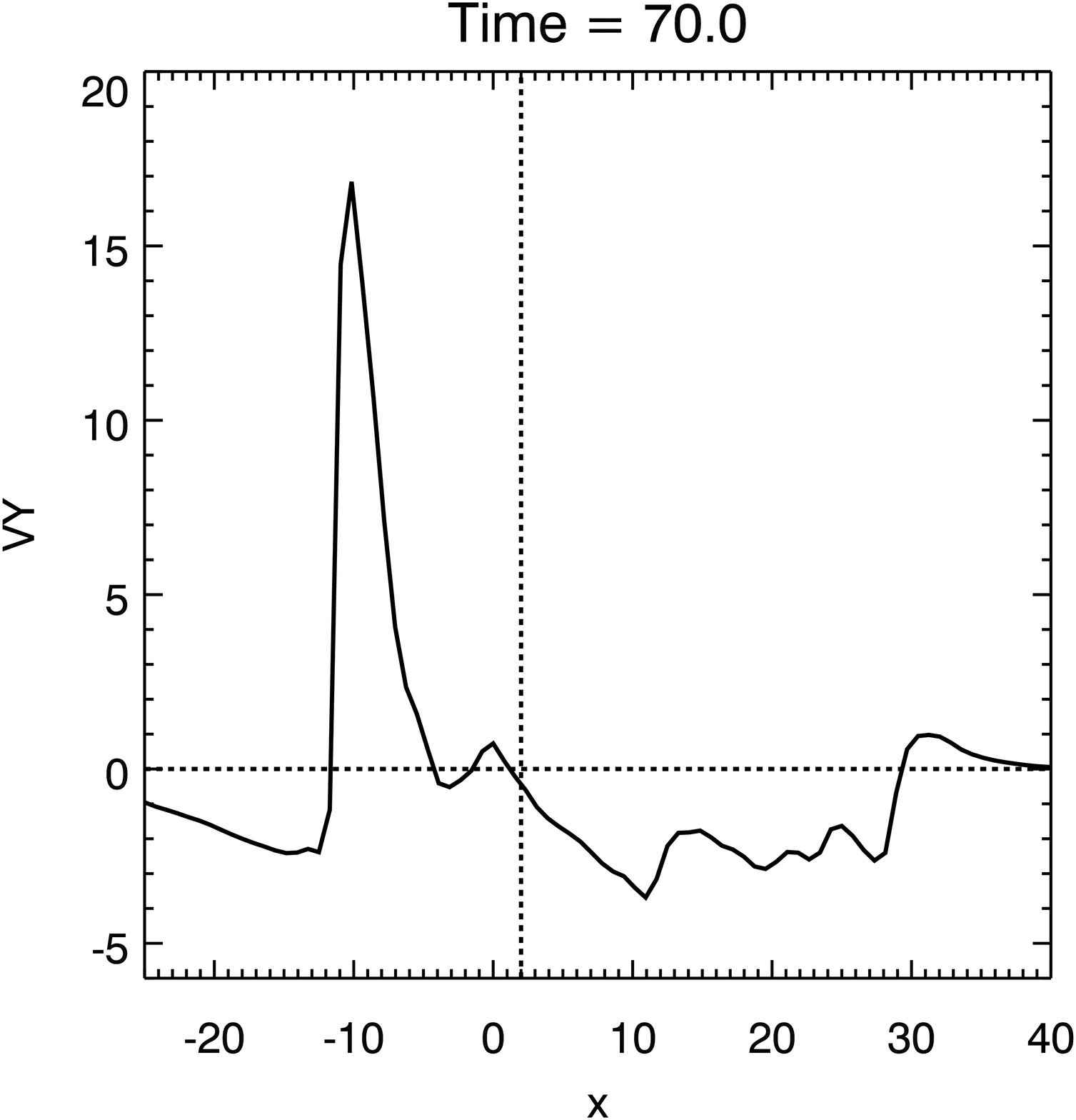}
   \caption{The upper left panel shows the magnetic reconnection rate using the ratio between
   the inflow and Alfv\'en speeds. Three vertical dashed lines in this panel outline 
   the times used in Figure \ref{fig03}. The other three panels are one dimensional distributions of density (RO),
   temperature (TE) and $y$-component velocity (VY) along the solid red line shown in 
   the lower panel of Figure \ref{fig03}. The horizontal dashed lines show the initial values while the vertical lines
   indicate the boundary between the hot and cold jets in these three panels. The density, length, temperature, time 
   and velocity units are $2.56 \times 10 ^ {-8}$ kg m$^{-3}$, $301.6$ km, $10000$ K, 
   $33.1 $ s and $9.09$ km s$^{-1}$, respectively.
   \label{fig04}}
\end{figure}

In this case, we set the parameters of the EMF to be $B_e = 32$ (corresponding
to $520$ G), $\eta_{max} = 0.1$ and $r_0 = 8$ (corresponding to
$\sim 2400$ km). The reconnection occurs when the local current density 
is larger than the threshold ($j_c = 10$) in the formula 
(\ref{Equ:Resistivity}). 
Figure \ref{fig03} depicts the results of the simulation at 
different times. In this figure the color stands for the
temperature, solid lines for magnetic field and arrows for velocity. The upper panel of 
Figure \ref{fig03} shows that the EMF already rose into the photosphere and chromosphere. 
At the time 55, the transition 
region has been pushed up to the height $y=35$ (i.e., about $10000$ km), and
fast reconnection has not happened yet. Later, when the magnetic reconnection
rate reaches the maximum (time=$70$, as 
shown by the upper left panel of Figure \ref{fig04}), a large 
amount of magnetic energy is released to the internal energy and kinetic energy. In this case, the 
reconnection occurs at the corona and the size of this microflare is about 50 from 
$x = -40$ to $10$ ($\sim 15000$ km, i.e. $\sim 20$ arcsec). 
The hot jet ($\sim 1.8 \times 10^6$K) and cold jet ($\sim 10^4$ K) have formed around this time that are similar to 
the results simulated by \cite{Yokoyama1995} and \cite{Nishizuka2008}. 

The reconnection rate and the one dimensional (1D) distributions of density, temperature and 
$y$-component of the velocity along the red line in Figure \ref{fig03} are illustrated in Figure \ref{fig04}. The reconnection 
rate is calculated by the ratio of the inflow speed ($V_{in}$) and the local Alfv\'en speed ($V_A$). Both two speeds are 
measured around the current sheet. The reconnection rate reaches the maximum around time=$70$ (38 minutes) and its value is about 0.09 that is within the value of 0.01 - 0.1 
predicted by \cite{Petshek1964}. The high reconnection rate partly results from 
some small plasmoids ejections from the X-point, 
which may increase the reconnection rate \citep{Shibata1995, Shibata1996}. While after time=$76$, some numerical
errors occur at the center of the EMF. Thus, the result after that time is unreliable although the reconnection rate increases again (see the upper left panel of Figure \ref{fig04}). A roughly estimated life time of this coronal microflare 
is about 12 minutes if we take the duration being from the beginning (time=$55$) to the ending (time=$75$) of 
the main reconnection. That is consistant with the observations values 
\citep[10 - 30 minutes;][]{Shimizu2002, Fang2006, Fang2010}. The other three panels show one dimensional 
distributions of density, temperature and $y$-component velocity at time=$45$, respectively. As we mentioned above, 
a hot jet and a cold surge are formed in this case. The hot jet ($\sim 1.8 \times 10^6$K) which originates 
from the reconnection region, is ejected to the higher corona with the velocity of $16$ 
(about 140 km s$^{-1}$), which corresponds to the observational EUV/SXR jet \citep{Chae1999, Brosius2009}. 
The denser cold surge ($\sim 10^4$ K and the density is two orders of magnitude larger than that of the hot jet), which is 
drawn to the corona by the hot jet. The cold surge falls to the lower atmosphere later with the speed of 3 
($\sim$ 30 km s$^{-1}$). This process is similar to the observational H$\alpha$/Ca surge \citep{Chae1999}.
The estimated 
size of the EUV/SXR or H$\alpha$ bright point in our simulation is about $50$ from $x = -40$ to $10$ 
($\sim 15000$ km, i.e. $\sim 20$ arcsec). 

\subsection{Chromospheric Microflare}
\label{Sec:Chromosphere_Case}

\begin{figure}[!htbp]
   \centering
   \includegraphics[width=300pt]{\figurepath/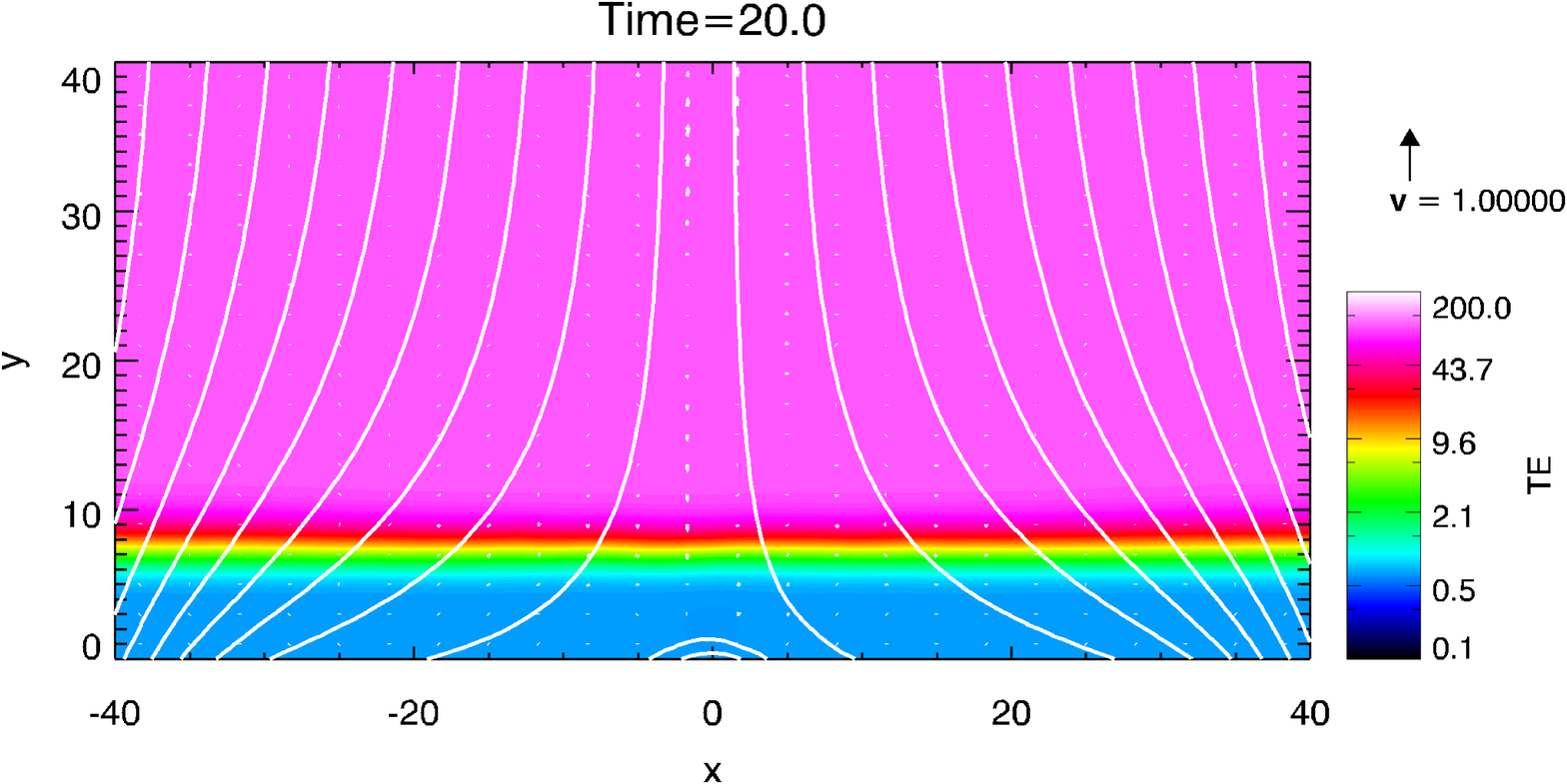}
   \includegraphics[width=300pt]{\figurepath/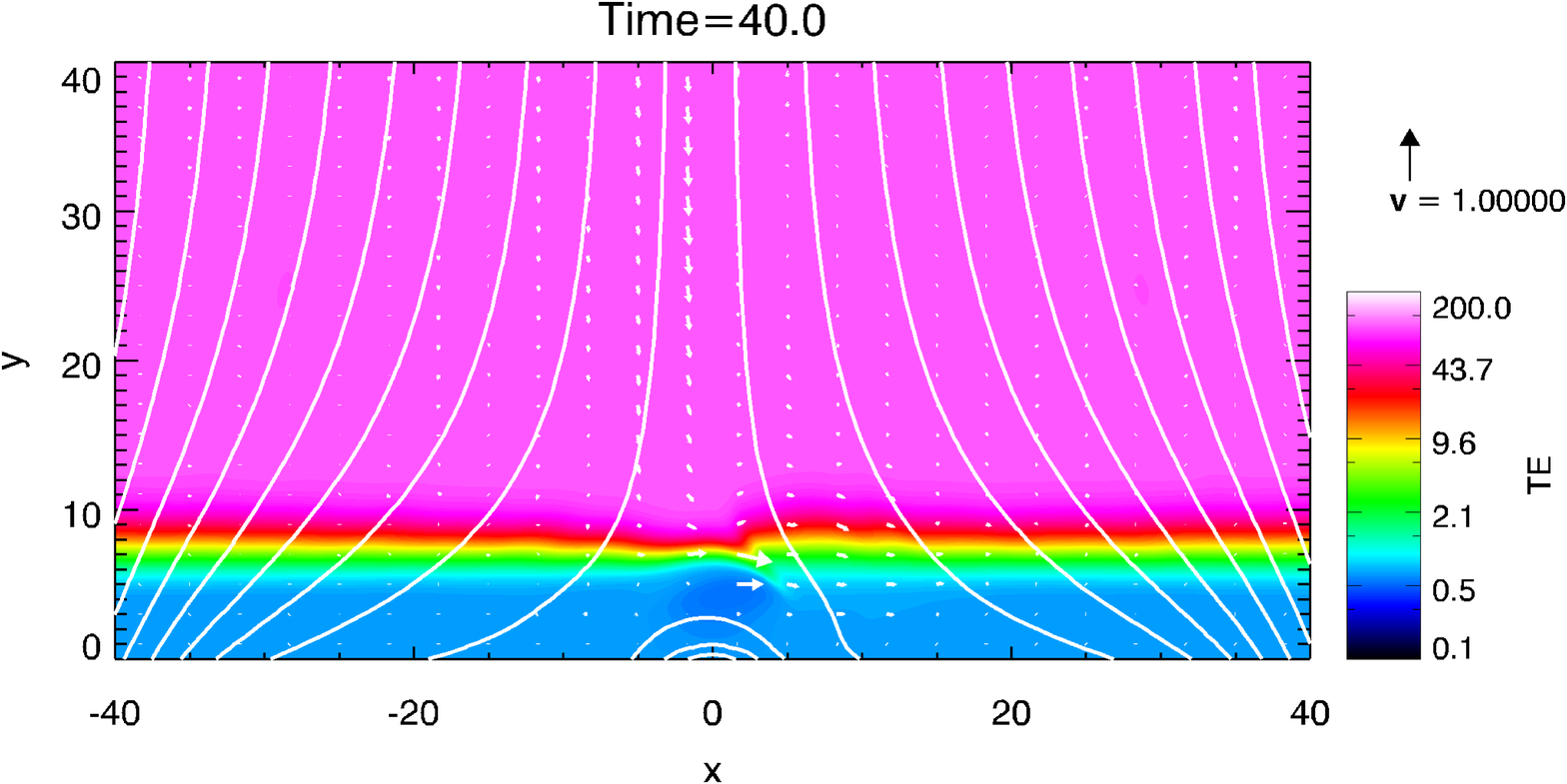}
   \includegraphics[width=300pt]{\figurepath/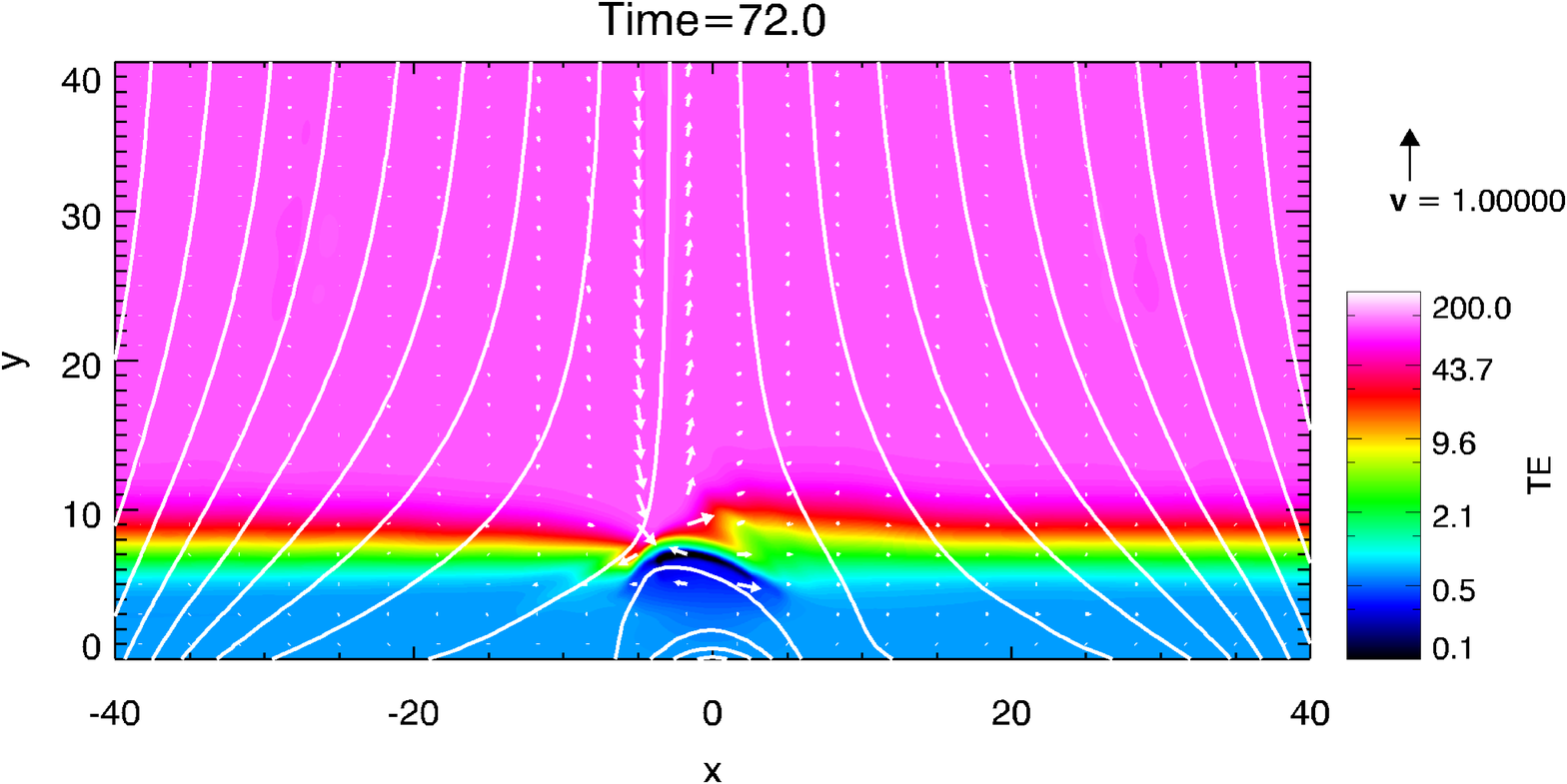}
   \caption{Temperature distributions (color scale), projected 
   magnetic field (solid lines) and velocity field (vector arrows) at
   time 20 (upper panel), 40 (middle panel) and 72 (lower panel).
   The units of length, temperature, time and velocity are $301.6$ km, 
   $10000$ K, $33.1 $ s and $9.09$ km s$^{-1}$, respectively.
   \label{fig05}}
\end{figure}

\begin{figure}[!htbp]
   \centering
   \includegraphics[width=200pt]{\figurepath/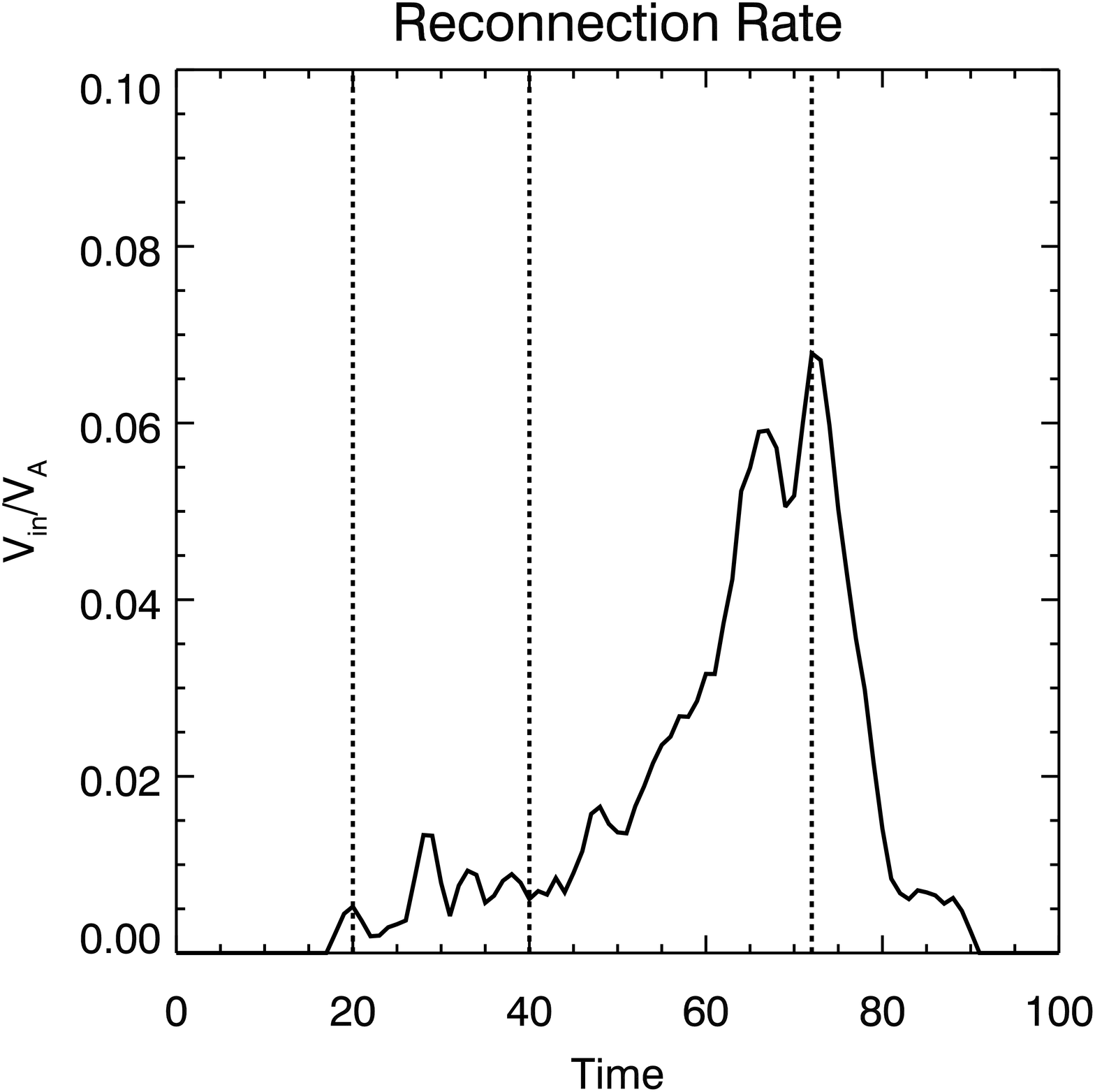}
   \includegraphics[width=200pt]{\figurepath/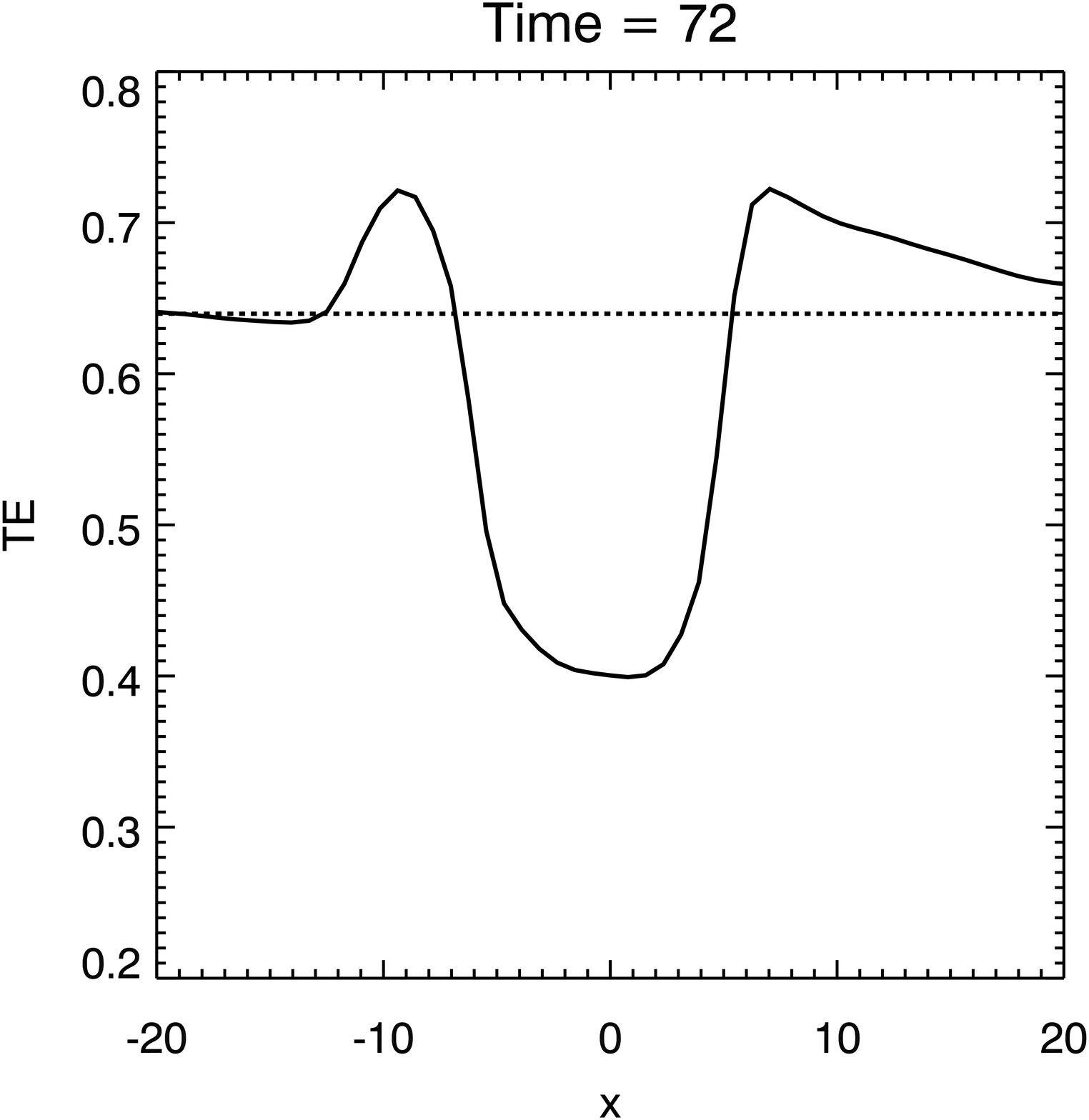}
   \caption{The left panel shows the magnetic reconnection rate using the ratio between
   the inflow and Alfv\'en speed. Three vertical dashed lines in this panel outline 
   the times used in Figure \ref{fig05}. The other panel are one dimensional distributions of 
   temperature (TE) along the solid red line shown in 
   the lower panel of Figure \ref{fig05}. The horizontal dashed lines show the initial value. 
   The length, temperature and time 
   units are $301.6$ km, $10000$ K and
   $33.1 $ s, respectively.
   \label{fig06}}
\end{figure}

In this case, the strength and half width of the EMF are set to be $B_e = 12$ ($200$ G), 
$\eta_{max} = 0.1$ and $r_0 = 6$ ($\sim 1800$ km). Compared the coronal case, the reconnection process 
is similar but the simulation result is totally differenct, since 
the coronal magnetic field prevents the weak EMF emerging to 
the height as high as the coronal one. Therefore, the magnetic field accumulated 
in the chromosphere until the local current density exceeds the threshold ($j_c=10$, the same as the coronal case). Finally, 
we got the chromospheric microflare resulting from a fast reconnection mainly 
in the chromosphere and partly in the transition region. 
The result of this chromospheric microflare is given by three snapshots in Figure \ref{fig05}. Similar to 
Figure \ref{fig03}, the three panels correspond to the times before the reconnection, the time when reconnection 
rate is increasing, and the time when the reconnection rate reaches to the maximum (as 
shown by the upper left panel of Figure \ref{fig06}), respectively. In this case, the 
reconnection occurs at the chromosphere and the size of this microflare is about 15 from 
$x = -10$ to $5$ ($\sim 4500$ km, i.e. $\sim 6$ arcsec). 
There is no obvious reconnection jets ejected into the corona and we only obtain 
some temperature enhancements in the chromosphere and partly in the transition region.

The reconnection rate and the one dimensional (1D) distribution of temperature along the red 
line in Figure \ref{fig05} are illustrated in Figure \ref{fig06}. 
The physical process is much simpler than the corona case. As the reconnection occurs in the 
dense chromosphere, it is very hard to heat the plasma to a very high temperature. 
The right panel of Figure \ref{fig06} is the temperature distribution at the 
the height of 1500 km. The inital value of the temperature is 0.64 (6400 K) and the temperature 
enhancement is about 800 K at this layer. The low temperature region is due to the expansion 
of the EMF. As we know, the plasma satisfies the \itshape frozen-in \upshape condition in the solar atmosphere 
without the resistivity. Therefore, the plasma of the lower atmosphere is dominanted by the strong magnetic 
field and finally such an adiabatic expansion will reduce the temperature. 
The low adiabatic index ($\gamma = 1.1$ in our simulations) can decrease such an effect, which also makes the 
simulated system more similar as an isothermal process. Although we can see two regions where the 
temperature is increased as seen in Figure \ref{fig06}, they are not be observed as two H$\alpha$ 
bright points since the hot region is a shell covering the EMF as seen from the lower panel of Figure \ref{fig05}. That is to say, we
can only observe one H$\alpha$ bright point if the line of sight is along 
the $y$-axis. The size of this bright point 
is about 15 from $x = -10$ to $5$ ($\sim 4500$ km, i.e. $\sim 6$ arcsec). There is no obvious temperature enhancement 
in the corona. Thus this case can explain the microflares with only H$\alpha$ emssion. The inflow (downward) 
speed in the lower panel of Figure \ref{fig05} is about 6 km s$^{-1}$ which may be observed as a redshifted Doppler velocity 
in coronal lines.
The outflow velocity of the reconnection flow in the chromosphere is about 20 km s$^{-1}$, but the direction of this 
outflow is redirected by 
background canopy-type field. Eventually, this coronal outflow ($\sim 4 $ km s$^{-1}$) moves 
upward next to the downward inflow as shown by the lower panel of Figure \ref{fig05}.

\subsection{Parameter Dependence}
\label{Sec:Parameter_Dependence}
So far we have simulated one coronal case and one chromospheric case. An
important question is which parameters determine whether reconnection happens
in the corona or in the chromosphere when an EMF appears. In this subsection, we study three parameters, 
i.e., the strength of the EMF ($B_e$), the maximum value of resistivity ($\eta_{max}$), and the half width of 
EMF ($r_0$). In order to understand the effects of different parameters, we perform extensive simulations
by changing one parameter with others being fixed.

\begin{figure}[!htbp]
   \centering
   \includegraphics[width=200pt]{\figurepath/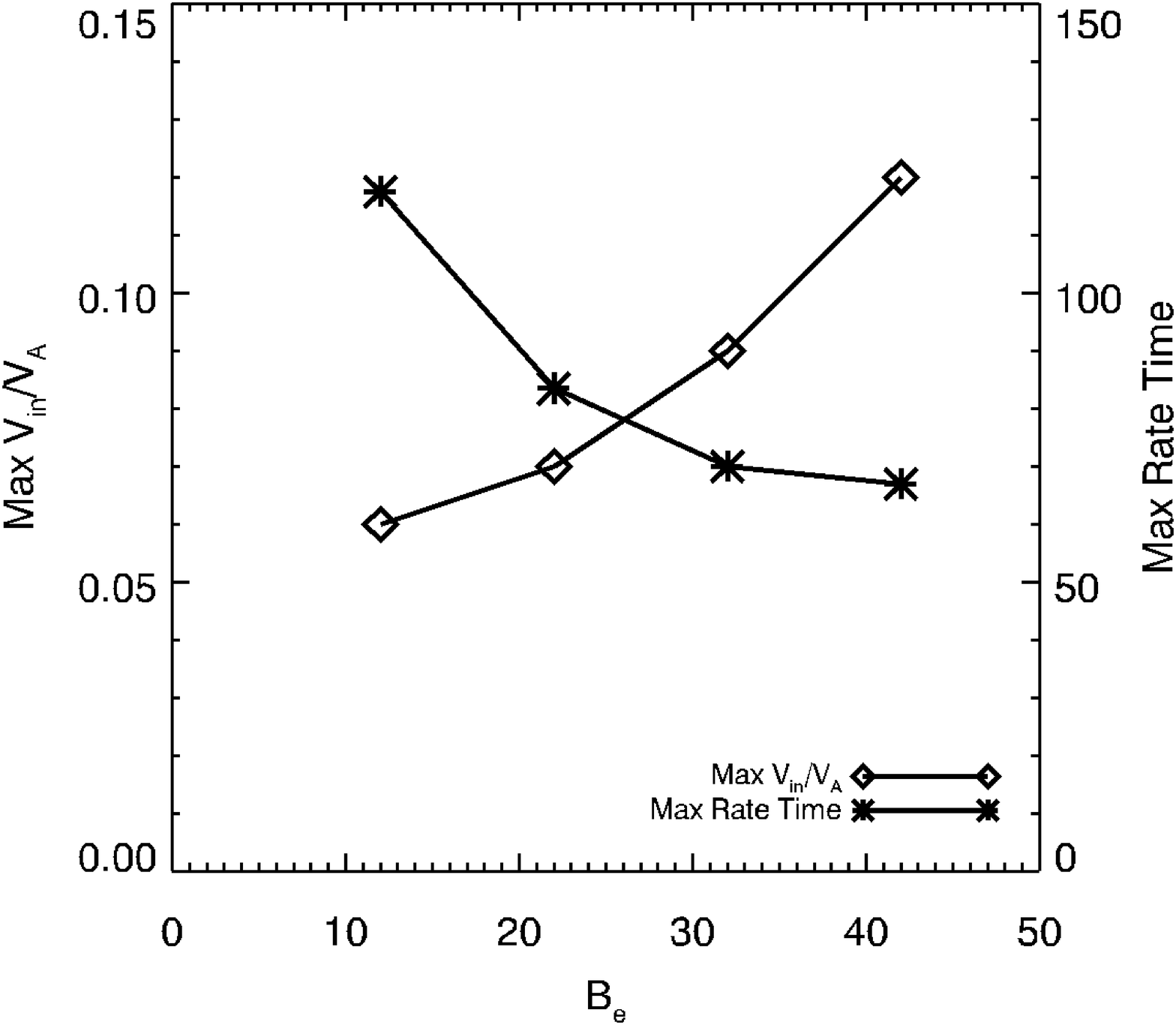}
   \includegraphics[width=200pt]{\figurepath/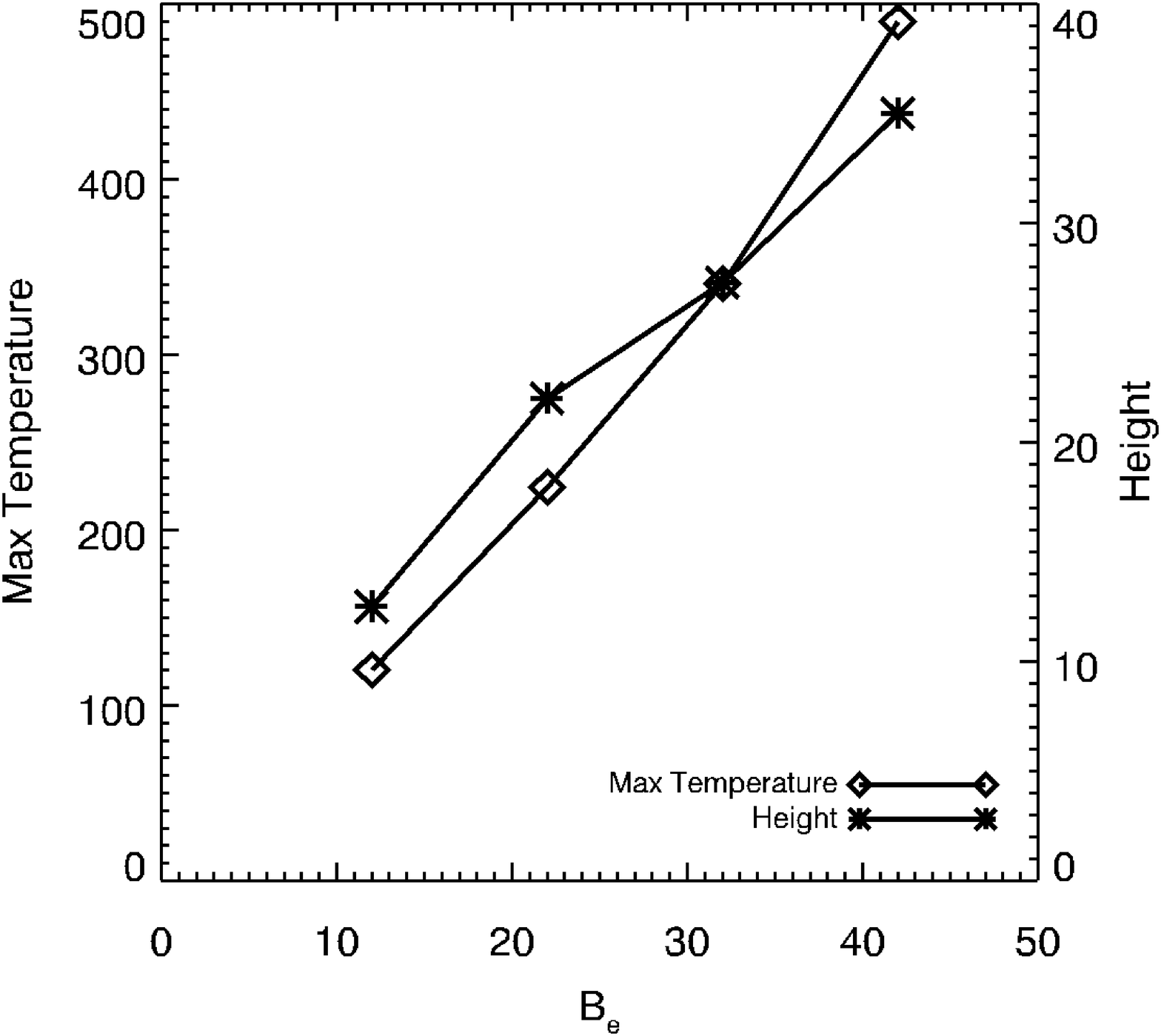}
   \caption{The left panel shows the dependence of \itshape Max \upshape $V_{in} / V_A$ (maximum magnetic reconnection rate) and 
   \itshape Max Rate Time \upshape (the time when the reconnection rate reaches 
   the maximum) on different strength of the EMF magnetic field.
   The right panel depicts the dependence of \itshape Max Temperature \upshape 
   (the maximum temperature at the \itshape Max Rate Time \upshape) and \itshape Height \upshape
   (the height of current sheet center at the \itshape Max Rate Time \upshape) on different strengths of EMF magnetic field.
   The length, temperature, time and magnetic field 
   units are $301.6$ km, $10000$ K, $33.1 $ s and
   $16.3$ G, respectively.
   \label{fig07}}
\end{figure}

Figure \ref{fig07} shows the dependence of the reconnection process on different strengths 
of the EMF magnetic field. Four values of $B_e$ are
studied, i.e., $B_e = 12, 22, 32$ and $ 42$, whereas the other two parameters are fixed, 
i.e., $\eta_{max} = 0.1$ and $r_0 = 8$. With the increasing of the strength of the EMF, the magnetic
reconnection rate can reach the maximum much faster and the maximum reconnection rate becomes larger 
and larger when we set a stronger initial magnetic field. The right panel shows how the maximum 
temperature and the height of the current sheet center change with the strength of EMF magnetic field. 
The \itshape Max Temperature \upshape and \itshape Height \upshape is very sensitive to the parameter $B_e$.
A stronger EMF can result to a higher altitude which can generate a stronger current density. Thus, 
it is easy to trigger the reconnection at an earlier time. The reconnection is faster and the outflow
is hotter than a weak EMF. When we set $B_e = 12$, we find that the \itshape height \upshape 
is 12.5 (3750 km), which is located nearly above the transition region. If we keep reducing the value $B_e$,
the reconnection will occur in the chromosphere.

\begin{figure}[!htbp]
   \centering
   \includegraphics[width=200pt]{\figurepath/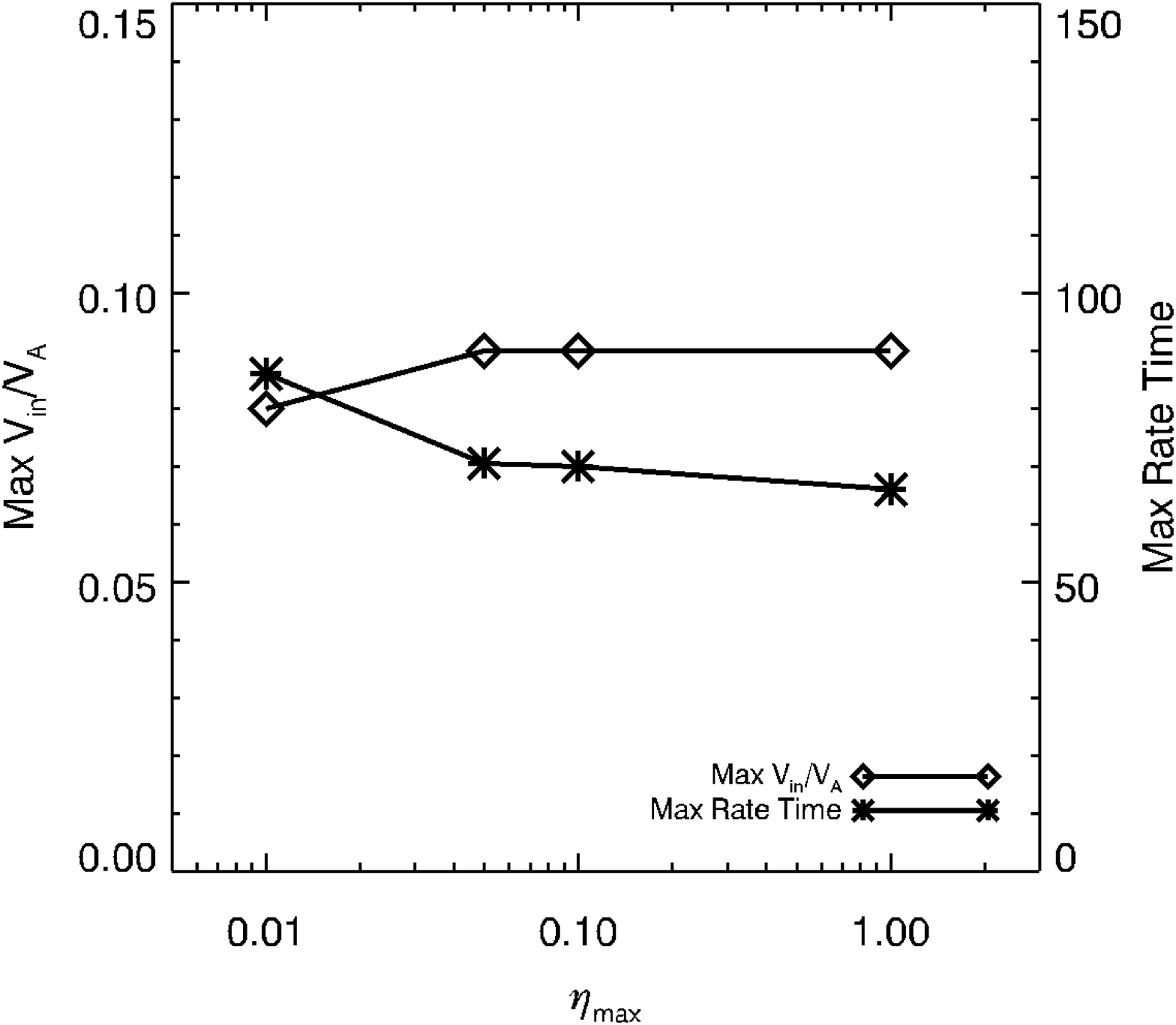}
   \includegraphics[width=200pt]{\figurepath/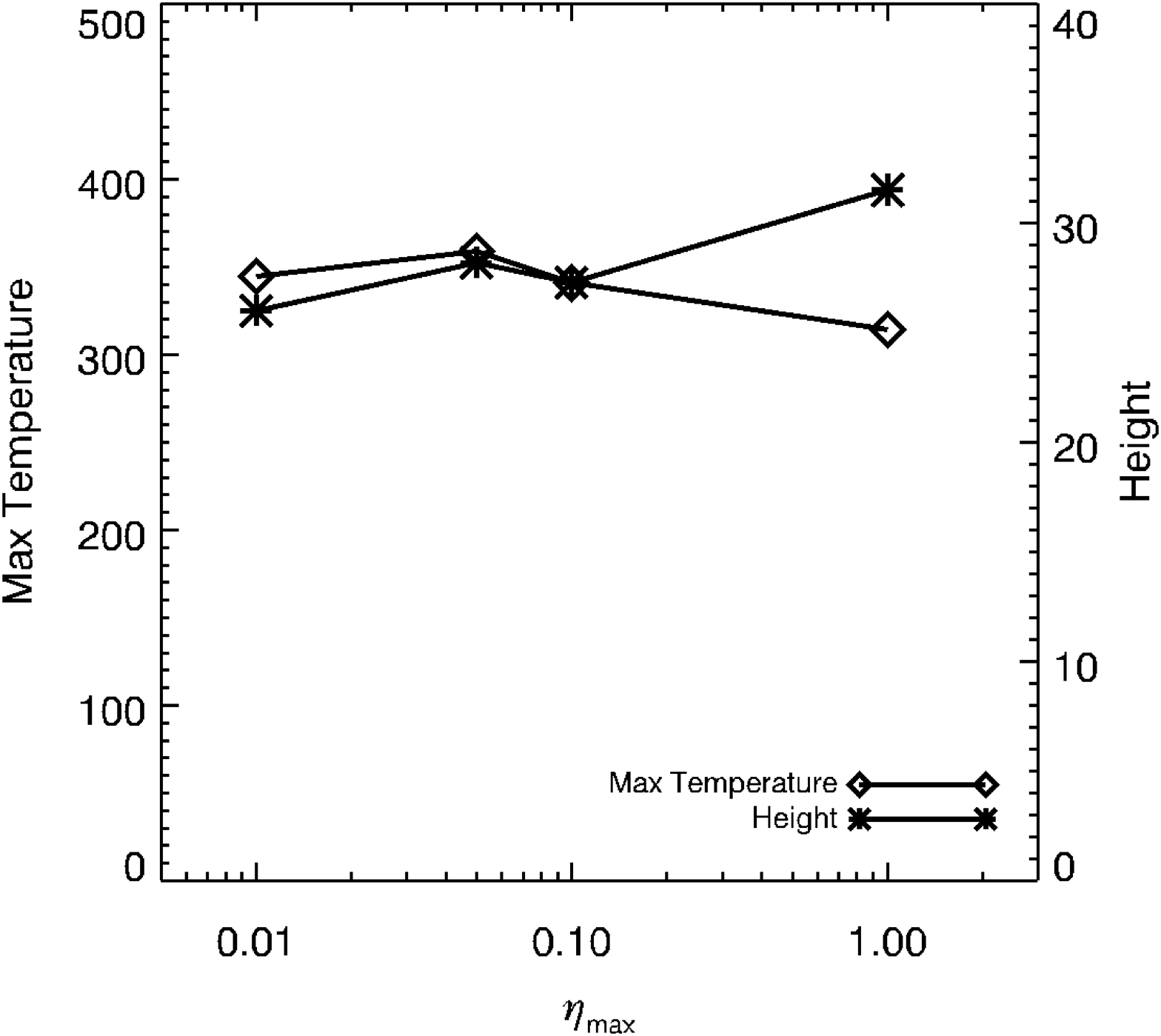}
   \caption{The left panel shows the dependence of \itshape Max \upshape $V_{in} / V_A$ and 
   \itshape Max Rate Time \upshape on different maximum values of resistivity ($\eta_{max}$).
   The right panel depicts the dependence of \itshape Max Temperature \upshape 
   and \itshape Height \upshape on different maximum values of resistivity ($\eta_{max}$).
   The length, temperature, time and resistivity 
   units are $301.6$ km, $10000$ K, $33.1 $ s and
   3445.1 $\Omega$ m, respectively.
   \label{fig08}}
\end{figure}

Figure \ref{fig08} shows the dependence of various quantities on different values of maximum resistivity ($\eta_{max}$). 
Four different values are selected, i.e., $\eta_{max} = 0.01, 0.05, 0.1$ and $ 1.0$. The other two parameters are fixed, 
i.e. $B_e = 32$ and $r_0 = 8$. We got the similar results as described in our previous paper \citep{Jiang2010}. 
The variation of the resistivity value does not change the results too much. The reconnections almost occur around
the heigh of 28 (8500 km) and the temperature enhancement at such a height is about 350 ($3.5 \times 10^6$). 

\begin{figure}[!htbp]
   \centering
   \includegraphics[width=200pt]{\figurepath/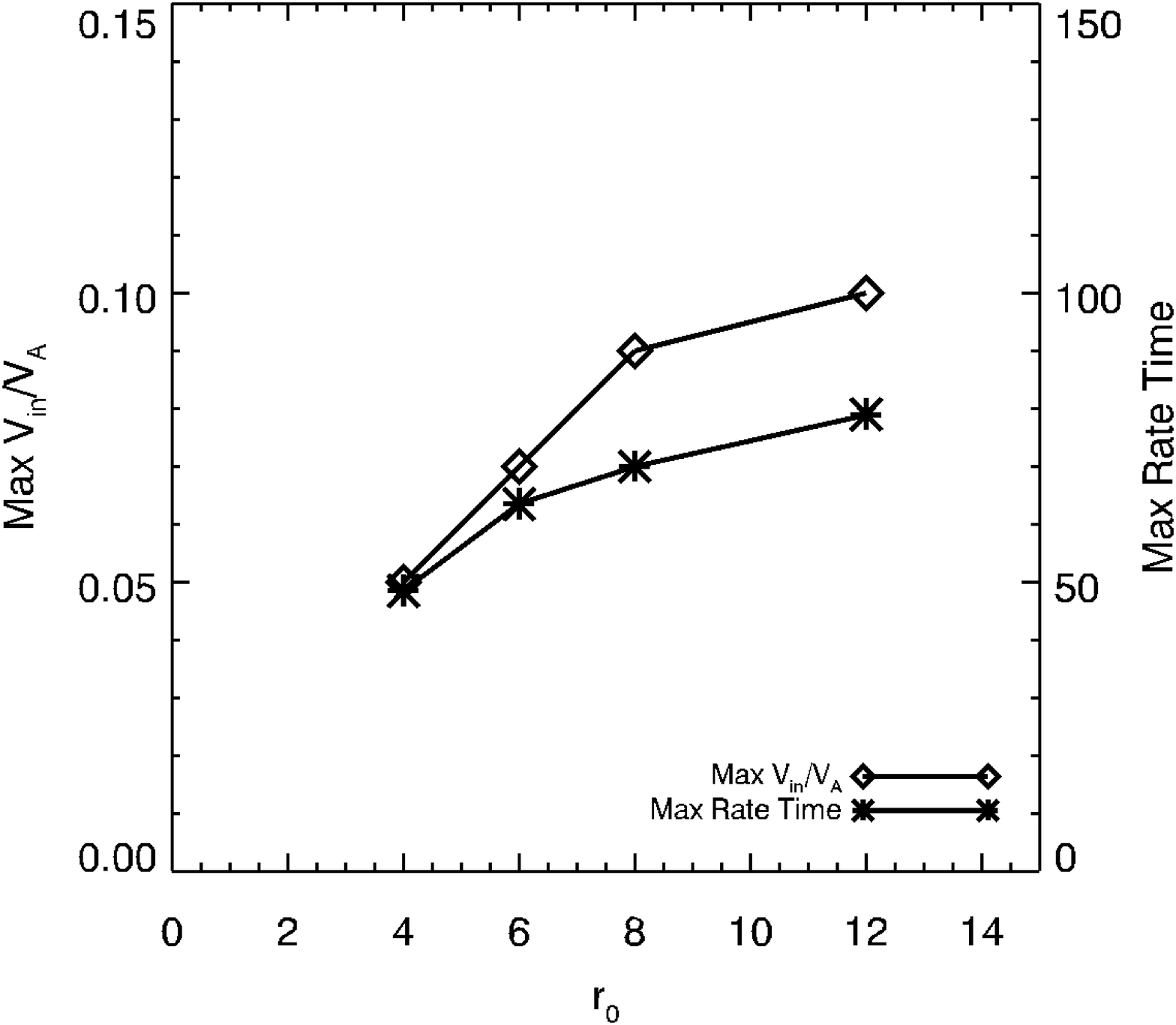}
   \includegraphics[width=200pt]{\figurepath/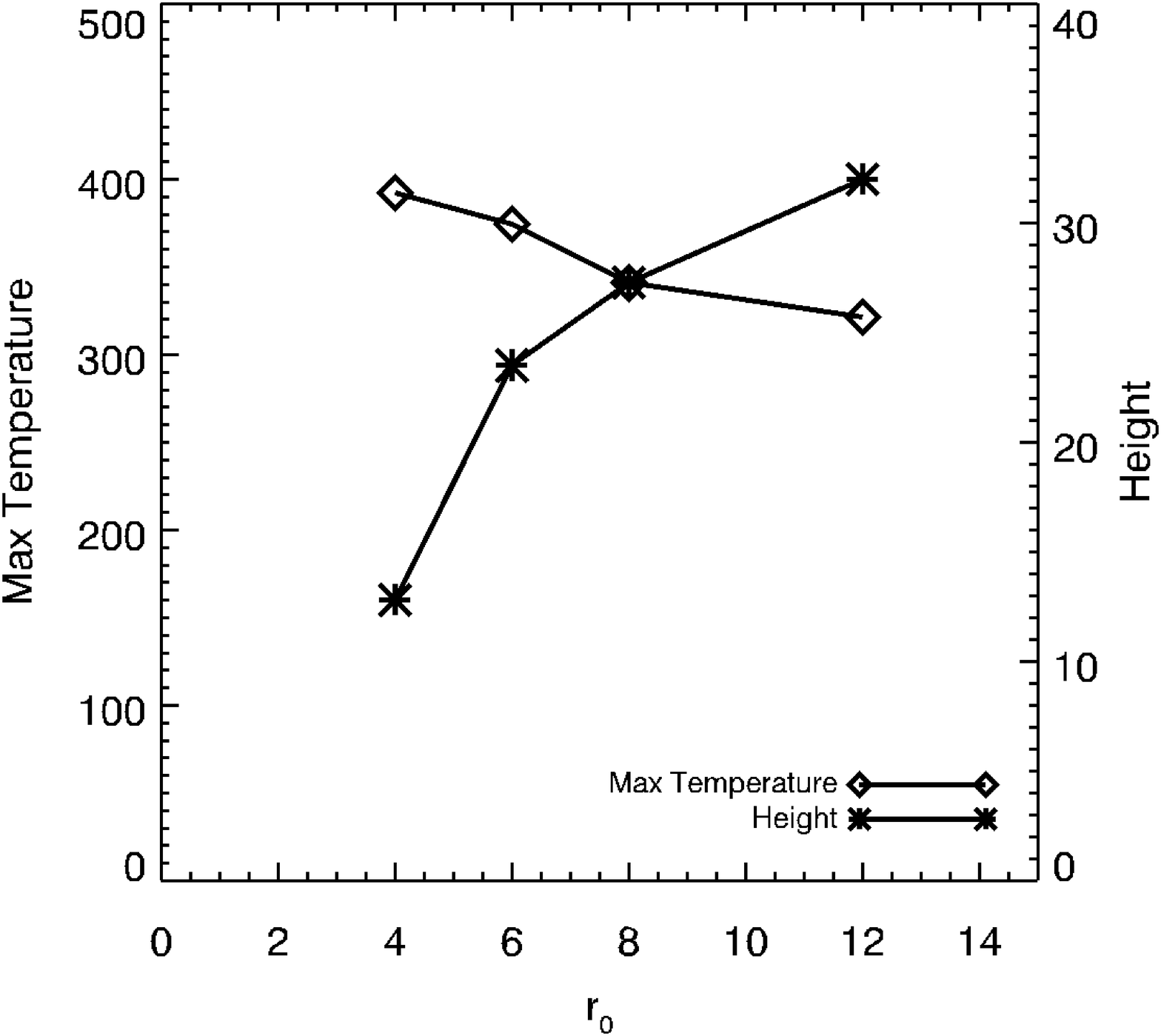}
   \caption{The left panel shows the dependence of \itshape Max \upshape $V_{in} / V_A$ and 
   \itshape Max Rate Time \upshape on different half width ($r_0$) of the EMF.
   The right panel depicts the dependence of \itshape Max Temperature \upshape 
   and \itshape Height \upshape on different half width ($r_0$) of the EMF.
   The length, temperature, time and length 
   units are $301.6$ km, $10000$ K, $33.1 $ s and
   $301.6$ km, respectively.
   \label{fig09}}
\end{figure}

The dependence on different half widths ($r_0 = 4, 6, 8$ and $12$) 
are depicted in Figure \ref{fig09}. The other two fixed parameters 
are $B_e = 32$ and $\eta_{max} = 0.1$. From the reconnection start 
time to the maximum time, a bigger EMF can rise to a higher altitude. 
We know that in the higher altitude the plasma $\beta$ is lower, 
thus the magnetic reconnection can be very fast with the same 
resistivity. At the same time, the strength of the magnetic field of 
EMF becomes weaker and weaker after expanding to such a long distance, 
so the temperature enhancement becomes smaller and smaller as indicated 
by Figure \ref{fig07}. Therefore, the left panel of Figure \ref{fig09} 
shows the reconnection rate and the reconnection duration increase with 
the increasing half width of the EMF. In the right panel, we find that 
the height of X-point is also sensitive to the size of the EMF. 
Combining the small magnetic strength of the EMF, we can get the result 
of chromospheric microflare as described in Section \ref{Sec:Chromosphere_Case}.

\section{DISCUSSION AND SUMMARY}
\label{Sec:Discussion_Summary}

\begin{figure}[!htbp]
   \centering
   \includegraphics[width=400pt]{\figurepath/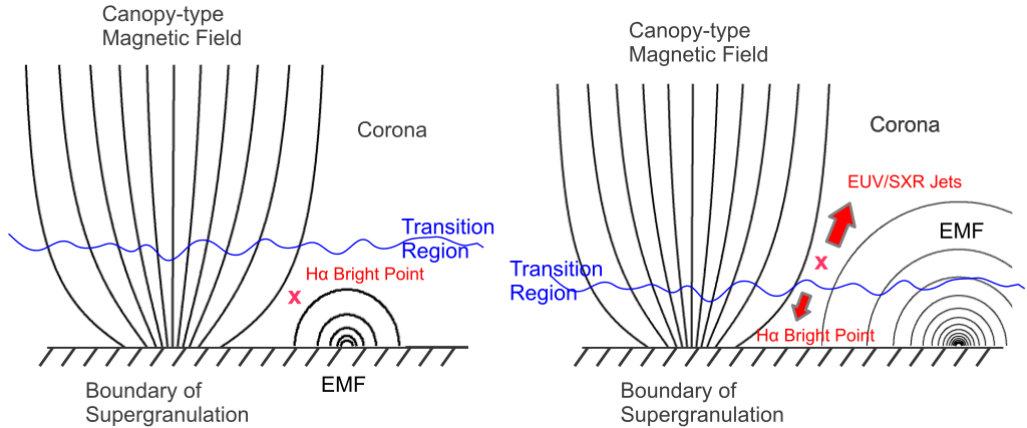}
   \caption{The cartoon shows the basic physical process of our simulated two 
   types of microflares, i.e. of chromospheric and coronal origins.
   \label{fig10}}
\end{figure}

In this paper, we simulate the microflares produced by EMFs. The EMFs with 
different strength and size lead to different results. Two types of the 
microflares in our simulations, i.e., of coronal origin and of chromospheric 
origin, can be illustrated by the cartoon in Figure \ref{fig10}. According 
to our simulations, small microflares with weak Doppler velocities and only 
H$\alpha$ emissions are most likely due to the reconnection in the 
lower solar atmosphere (for instance, the chromosphere or photosphere). 
However, big microflares with obvious Doppler velocities and emissions in 
both low and high temperature wavelengthes (H$\alpha$ and EUV/SXR) 
originate from the reconnection in the corona. This model is similar to the
unified model of \citet{Shibata1996, Shibata2007} and \citet{Chen1999} in 
the sense that the different behaviors of eruptions are determined by the 
different height of the magnetic reconnection. According to the result in 
this paper, the strength and the size of the EMF determines whether the 
microflare is of coronal origin or of chromospheric origin when the EMF 
emerges from the center of a supergranule, probably dragged by the convective 
upflow. Of course, EMFs may also appear at other sites of a supergranule, 
which would also affect the height of the reconnection site.

\begin{figure}[!htbp]
   \centering
   \includegraphics[width=300pt]{\figurepath/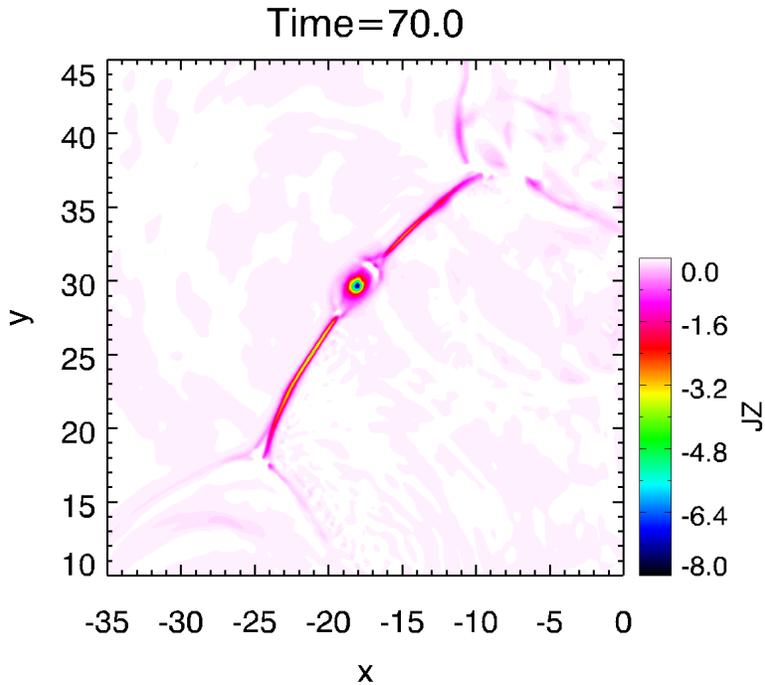}
   \caption{A amplified view of $z$-component current density (JZ) distribution 
      of the conoral case at the time 70. The overall view of 
      temperature distribution is shown by the bottom panel of Figure 
      \ref{fig03} at the same time. The range of this box is [-35, 0] 
      $\times$ [10, 45], in which a plasmoid in the center of the current 
      sheet is clearly seen.
   \label{fig11}}
\end{figure}

As mentioned in Section \ref{Sec:Corona_Case}, we find some plasmoids in our 
simulations. Figure \ref{fig11} shows the current density distribution of 
a small region [-35, 0] $\times$ [10, 45] in Figure \ref{fig03}. 
From both the temperature and current density distributions (Figure 
\ref{fig03} and \ref{fig11}), we can see a plasmoid is formed at the center 
of the current sheet. The ejection of the plasmoid can increase the 
magnetic reconnection rate temporarily \citep{Shibata1995, Shibata1996, Shen2011}. 
In the reconnection process, the plasmoids are generated one by one, 
therefore we can see some oscillations of the reconnection rate in Figure 
\ref{fig04}. However, for the chromospheric case, the Alfv\'en speed is much 
smaller than the coronal one, which leads to a smaller magnetic Reynolds number. 
Thus, similar to the results presented by \cite{Shen2011}, there is no plasmoid 
produced in the chromospheric case. The corresponding reconnection rate in 
Figure \ref{fig06} also shows less oscillations. In the coronal case, as the 
reconnection occurs in the corona, both hot jets
and cold surges are ejected. No doubt that the EUV/SXR emissions from the hot 
jet will be observed first, whereas H$\alpha$/Ca bright points and surges 
appear later. If we take the temperature response at the height of 1500 km, 
which is the H$\alpha$ formation height \citep{Vernazza1981}, the roughly 
calculated time delay between the EUV/SXR and H$\alpha$ emissions is about 
3--5 minutes, which is comparable to the observations \citep{Zhang2012}. We 
have to mention that the temperature enhancement and the time delay can be 
more realistic if we include the thermal conduction, radiation, the high 
energe particles (non-thermal electrons and ions), and the effect of partially 
ionised plasma. For simplicity and for the sake to see how emerging
flux with different sizes and strengths influencesthe reconnection, we kept 
the threshold of the current density $j_c$ in the anomalous resisitivity in 
Eq. (\ref{Equ:Resistivity}) to be constant. In reality, the threshold of $j_c$ in the
chromosphere might be higher because of its high plasma density. In order to 
see how $j_c$ would change the reconnection process, we performed other numerical
experiments with higher $j_c$ in the chromosphere, and it is found that the basic results are the 
same, the main difference is that the commencement of reconnection is postponed 
only for the case that the current sheet located at the chromosphere,
as demonstrated by \citet{Yokoyama1994}.

In summary, we successfully simulated the coronal and chromospheric microflares 
by using MHD simulations with a canopy-type magnteic configuration. The 
different observational behaviors between coronal microflares and chromospheric 
ones are due to the height of magnetic reconnection, which is determined by the
size of the emerging flux, i.e., smaller emerging flux produces chromospheric
microflares and larger emerging flux produces coronal microflares.
Correspondingly, the resulting microflares have different sizes. In the case 
of coronal origin, the sizes of the simulated microflares are $11000-16000$ km, 
i.e., $15-22$\arcsec, which correspond to big microflares. We find a hot 
jet ($\sim$$1.8 \times 10^6$ K for the typical case), which should be relates to the observational 
EUV/SXR jet, and a cold jet ($\sim$$10^4$ K for the typical case), which corresponds to the 
observational H$\alpha$/Ca surge or brightening. Some plasmoids generated at 
the center of the current sheet are ejected, which can increase the magnetic 
reconnection rate. In the case of chromospheric origin, the sizes of the 
microflares are $4200-4500$ km, i.e., $\sim$6$\arcsec$, which correspond to 
relatively small microflares. As the reconnection occurs in the chromosphere, 
only the H$\alpha$/Ca brightenings show up in this case, where no significant 
SXR brightening and no plasmoids are produced in this case. The parameter 
survey qualitatively shows that the size and strength of the EMF are the key 
parameters which determine the height of the reconnection X-point. For some 
typical values, we can get the reconnection either in the corona or in the 
chromosphere. The parameter $\eta_{max}$ has little effect on the final results. 

This paper deals with the microflares related to EMFs. It should be mentioned
that some microflares are due to other mechanisms, e.g., the reconnection
between the MMFs \citep{Harvey1973} and the pre-existing magnetic field. 
\citet{Priest1994} proposed a canceling magnetic features model. 
\citet{Browning2008} performed 3D MHD simulations to study the role of kink 
instability in the magnetic energy releas process, which may also lead to
microflares. It is also noticed that the simulated emerging flux in this paper
does not include the twisted field, which is required for the emerging flux
to survive the subsurface convection \citep[see][and references
therein]{Babcock1961, Piddington1975, Fan2009}, e.g., \cite{Murray2008} found that the tension force of
twisted tube plays a key role in determining whether the flux can successfully
emerge to the upper solar atmosphere in 3D numerical experiments. In the
2D coronal simulations, the twisted field can be easily considered by including
$B_z$, which would not change the results greatly. Moreover, a single-fluid
model for the solar atmosphere was adopted in this paper for simplicity. In
reality, the chromosphere, especially the lower part, is weakly-ionzed,
and partial ionization should be considered, which would lead to strong
anisotropic Cowling resistivity \citep{Cowling1957, Khodachenko2004, Leake2006}, ambipolar diffusion 
\citep{Vishniac1999, Soler2009, Singh2011}. Moreover, the ionization of neutral atoms would consume
a significant part of the released energy in the magnetic reconnection
\citep{Chen2001, Jiang2010}.

\acknowledgments
The authors are grateful to the referee for suggestive comments and to C. 
Xia and F. Chen for helpful discussions. The computations were done by using 
the IBM Blade Center HS22 Cluster at High Performance Computing Center (HPCC) 
of Nanjing University of China. This work is supported by the National Natural
Science Foundation of China (NSFC) under the grants 10221001,
10878002, 10403003, 10620150099, 10610099, 10933003, 11025314, and 10673004, as
well as the grant from the 973 project 2011CB811402.

\clearpage

\end{document}